\documentclass[11pt,draftcls,onecolumn]{IEEEtran}
\usepackage{graphicx}
\usepackage{amsmath,amssymb,amsfonts,algorithm,algorithmic,epsfig,subfigure,setspace,epstopdf}
\usepackage[finalnew]{trackchanges}%,dsfont}
\usepackage[center]{caption}

\long\def\comment#1{}

\newfont{\bbb}{msbm10 scaled 700}

\newfont{\bb}{msbm10 scaled 1100}

\newcommand{\RR}{\mbox{\bb R}}

\newcommand{\fv}{{\bf f}}

\newcommand{\uv}{{\bf u}}

\newcommand{\xv}{{\bf x}}
\newcommand{\yv}{{\bf y}}

\newcommand{\zerov}{{\bf 0}}

\newcommand{\Am}{{\bf A}}

\newcommand{\Dm}{{\bf D}}

\newcommand{\Fm}{{\bf F}}
\newcommand{\Gm}{{\bf G}}
\newcommand{\Hm}{{\bf H}}
\newcommand{\Id}{{\bf I}}
\newcommand{\Jm}{{\bf J}}

\newcommand{\Lm}{{\bf L}}

\newcommand{\Pm}{{\bf P}}

\newcommand{\Sm}{{\bf S}}
\newcommand{\Tm}{{\bf T}}
\newcommand{\Um}{{\bf U}}

\newcommand{\Bc}{{\cal B}}

\newcommand{\Ec}{{\cal E}}

\newcommand{\Nc}{{\cal N}}
\newcommand{\Oc}{{\cal O}}

\newcommand{\Sc}{{\cal S}}

\newcommand{\Vc}{{\cal V}}

\newcommand{\Lcb}{\bf {\cal L}}

\newcommand{\betav}{\hbox{\boldmath$\beta$}}

\addeditor{SN}
\title{Perfect Reconstruction Two-Channel Wavelet Filter-Banks for Graph Structured Data}
\author{Sunil~K.~Narang*,~\IEEEmembership{Student~Member,~IEEE,}
        and~Antonio~Ortega,~\IEEEmembership{Fellow,~IEEE}% <-this % stops a space
\thanks{Sunil K. Narang is with the Signal \& Image Processing Institute, Ming Hsieh Department of Electrical
Engineering, University of Southern California, Los Angeles, California, USA 90089 email: narang.sunil@gmail.com.}
\thanks{Antonio Ortega is with the Signal \& Image Processing Institute, Ming Hsieh Department of Electrical
Engineering, University of Southern California, Los Angeles, California, USA 90089 email: antonio.ortega@sipi.usc.edu.}
\thanks{This work was supported in part by NSF under grant CCF-1018977.}}
\newtheorem{lemma}{Lemma}

\newtheorem{proposition}{Proposition}

\begin{document}

\maketitle
% \pagenumbering{arabic}
\begin{abstract}
In this work we propose the construction of two-channel 
wavelet filterbanks
% perfect reconstruction wavelet-like filterbanks 
for analyzing functions defined on the vertices of any
arbitrary finite weighted undirected graph. These graph based functions 
are referred to as {\it graph-signals} as we build a framework 
in which many concepts from the classical signal processing domain, 
such as Fourier decomposition, signal filtering and downsampling
can be extended to graph domain. Especially, we observe 
a {\em spectral folding} phenomenon in {\em bipartite graphs} which occurs 
during downsampling of these graphs and produces {\em aliasing} in graph signals.
\add[SN]{This property of bipartite graphs, allows us to design critically 
sampled two-channel filterbanks, and}
we propose quadrature mirror filters 
(referred to as graph-QMF) for bipartite graph which cancel aliasing 
and lead to perfect reconstruction. For arbitrary graphs we present a bipartite subgraph 
decomposition which produces an edge-disjoint collection of bipartite 
subgraphs. Graph-QMFs are then constructed on each bipartite subgraph leading 
to ``multi-dimensional'' separable wavelet filterbanks on graphs. Our proposed 
filterbanks are critically sampled and we state necessary and sufficient 
conditions for orthogonality, 
aliasing cancellation and perfect reconstruction. 
The filterbanks 
are realized 
by Chebychev polynomial approximations. \\
{\bf Note:} Code examples from this paper are available at 
http://biron.usc.edu/wiki/index.php/Graph\_Filterbanks

\end{abstract}
% \begin{IEEEkeywords}
% Wavelets on Graphs, Graph Signals \& Graph Filterbanks, Downsampling in Graphs
% \end{IEEEkeywords}
\begin{center} \bfseries EDICS Category: DSP-WAVL, DSP-BANK, DSP-MULT, DSP-APPL, MLT \end{center}
\IEEEpeerreviewmaketitle
\section{Introduction}
\label{sec:intro}
\subsection{Motivation}
Graphs provide a very flexible model for 
representing data in many domains. Many networks such 
as biological networks~\cite{Weber'05}, 
social networks~\cite{Crovella'03,Newman'02} and sensor 
networks~\cite{Godwin'08,Ramchandran'06} etc.
have a natural interpretation in 
terms of finite graphs with 
vertices as data-sources and links  
established based on connectivity, 
similarity, ties etc.  
%SN: Graph Signal Interpretation
The data on these graphs 
can 
% thus 
be visualized as a finite 
collection
of samples termed as {\em graph-signals}. 
For example, graphical models can be 
used to represent 
irregularly sampled datasets in 
Euclidean spaces \change[SN]{ or  missing samples in regular 
grids}{such as regular grids with missing samples}. 
% Whereas there is a vast literature on signal processing tools designed 
% for data defined on regular Euclidean spaces, the development of 
% methods that are able to accommodate complicated data domains is an
% important problem.
In many machine learning applications 
multi-dimensional datasets can be 
represented as {\em point-clouds} of vectors 
and links are 
established between data sources based on the 
distance between their feature-vectors. 
% In this context neighborhood preserving 
% dimensionality reduction techniques such as 
% ~\cite{NPPCA} for high-dimensional 
% dataare mentionable. 
% SN: consider neighbor preserving  dimensionality reduction of high-dimensional data
In computer vision, 
meshes are polygon 
graphs in 2D/3D space and the attributes of 
the sampled points 
(coordinates, intensity etc) constitute 
the graph-signals. 
The graph-signal 
formulation can also 
be used to solve systems  
of partial differential equations 
using finite element 
analysis (grid based solution).
\change[SN]{Mostly, t}{T}he
sizes (number of nodes) of the graphs 
in these applications \change[SN]{are enormous}{can be very large}, 
% (imagine millions of facebook users),
%  or 
% thousands of protein molecules 
% interacting in a human body), 
which present computational 
and technical challenges 
% are unmanageable 
for the purpose of 
% in 
storage, analysis etc.  
In some other applications 
% domains 
such as wireless sensor-networks,  
the data-exchanges between far-off 
nodes can be expensive 
(bandwidth, latency, energy constraints issues). 
Therefore, instead of operating 
on the original graph, 
it would be desirable to find and operate on
smaller graphs with 
fewer nodes and data 
representing a smooth\footnote{more generally, it could be any sparse approximation of the original data.} 
approximation of 
the original data. 
Moreover, 
% for scalability, 
such 
% algorithms 
systems 
% would 
need to 
employ  localized 
operations which could be 
computed at each node by 
using data from 
a small neighborhood 
of nodes around it. 
% scalability of such methods, the approximation 
% to develop scalable algorithms 
% which employ 
% % . Another approach, 
% % would be 
% % to 
% and to provide a 
% coarse approximation of the graph with fewer nodes and 
% data representing a smooth approximation of original data. 
Multi-channel wavelet filterbanks,
widely used as a signal processing tool for 
the sparse representation of signals, possess both 
these features (i.e. smooth approximations and localized operations).
% (downsampling and localized operations). 
% Especially 
For example, a two channel wavelet transform 
splits the sample 
space into an approximation subspace which 
contains
a smoother (coarser) version of 
the original signal
and a detail subspace containing additional 
details required to perfectly 
reconstruct the original signal.
%  . 
% The approximation subspace  and the detail 
% subspace 
% a two channel 
% transform splits the sample 
% space into an approximation subspace
% and a detail subspace. 
% The approximation subspace contains
% a smoother (coarser) version of 
% the original signal and the detail 
% subspace contains additional 
% details required to perfectly 
% reconstruct the original signal. 
% These filters are locally 
% supported in the spatial 
% (temporal) domain of the signal which means each filter 
% can be computed around a sample $n$ by using data from 
% a small neighborhood of samples around $n$. Moreover the 
% filters are localized in the frequency domain, 
% are critically sampled 
% and provide perfect reconstruction (no loss in transformation).
% as well as 
% in frequency domain, 
% are invertible and critically sampled. 
% localized support on the signal 
A discussion of the construction 
and analysis of 
% these 
% two-channel 
wavelet filterbanks 
for regular signals can be
found in standard textbooks 
such as \cite{Vetterli_book}. 
While wavelet transform-based 
techniques would 
seem well suited
to provide efficient local analysis, 
a major obstacle to their
application to graphs is that these, 
unlike images, are not
regularly structured. 
For graphs traditional notions of 
dimensions along which to filter
the data do not hold.
% Moreover, 
% classical wavelet transforms are 
% critically sampled as they use 
% local filtering operations 
% followed by downsampling. In
% a graph, there is no obvious way to downsample nodes
% in a regular manner, since the neighboring nodes 
% vary in
% number and orientation. Therefore a different framework 
% is needed
% to study graph based transforms.
% \subsection{Existing Work}
% In general, the transform techniques
% in graph analysis literature
% can be broadly divided into a) global methods,
% e.g., those using concepts of graph spectral theory, 
% and b) wavelet like localized methods which are supported 
% on a local neighborhood around each node. 
% Global methods are often 
% based on the Laplacian matrix, 
% whose eigenvalues and 
% eigenvectors contain 
% global information 
% about the shape of 
% the graph.  
% Common 
% applications of global methods 
% include, graph partitioning (graph-cuts)~\cite{Newman'02}, simplification, 
% graph based 
% feature extraction~\cite{kondor} 
% and graph matching
% \cite{hancock}.  
% A comprehensive discussion of global
% methods can be found in
% \cite{fchung} and
% \cite{luxberg'07}. 
% In addition to uncovering mostly 
% global information, 
% global methods do not scale 
% well as the graph size increases,
% e.g., the time 
% required to perform the 
% eigenvalue decomposition 
% can be significant. %SN2 : So we need something else
% Therefore 

Researchers 
have recently focused on 
developing 
localized transforms specifically for data 
defined on graphs. 
Crovella and Kolaczyk ~\cite{Crovella'03} designed
wavelet like functions on graphs 
which are localized in space and time. 
These graph functions $\psi_{j,k}$ are 
composed of either shifts or dilations 
of a single generating function $\psi$.
Wang and Ramchandran ~\cite{Ramchandran'06} 
%who 
proposed graph dependent basis
functions for sensor network graphs, 
which implement an invertible $2$-channel like 
filter-bank.  
There exists a natural spectral interpretation 
of graph-signals in terms of eigen-functions 
and eigen-values of graph 
Laplacian matrix $\Lm$. 
Maggioni and Coifman~\cite{Coifman'06} 
introduced ``diffusion wavelets''
as the
localized basis functions of the 
eigenspaces of the dyadic powers of 
a diffusion
operator.
Hammond et. al.~\cite{Hammond'09} 
construct a class of wavelet operators 
in the {\it graph spectral domain}, 
i.e., the space of 
eigenfunctions of the 
graph Laplacian matrix $\Lm$. 
These \change[SN]{basis functions}{eigenfunctions} provide a spectral 
decomposition 
for data on a graph similar 
to the Fourier transform 
for standard signals. A common drawback of
all of these filterbank designs is 
that they are not critically sampled \change[SN]{ 
% and
as   
% 
% do not 
% address downsampling in graph. As a result 
}{:}
the output of the transform is not downsampled and there is 
oversampling 
\remove[SN]{usually} by a factor \change[SN]{of}{equal to the} number of 
channels in the filterbank. Unlike  
classical wavelet transforms which have 
well-understood 
downsampling/upsampling operations, 
% followed by downsampling. In
there is no obvious way in graphs to downsample nodes
in a regular manner, 
since the neighboring nodes 
vary in
number\remove[SN]{ and orientation}.
% This can cause scalability problems for 
% large graphs and hinders the use 
% of graph based 
% transforms in compression 
% and high-dimensional data-analysis
% % denoising 
% applications.
Lifting based wavelet transforms have been 
proposed in~\cite{Wagner'06,Silverman}
for graphs in Euclidean Space and in our previous work  
for trees in~\cite{Godwin'08,GodwinJ} and for general 
graphs in \cite{Narang:APSIPA'09}. 
These transforms are critically sampled and 
invertible by construction. However the design requires 
splitting the vertex set of the graph into two disjoint 
sets and \add[SN]{the} transform
is computed only on the links 
between nodes in different sets. 
Thus links between nodes in same set are not utilized 
by the transform. 
% and are lost. 

Our \remove[SN]{prime} contribution in this paper 
is 
to introduce 
% the introduction of  
% {\em for the first time},  
\change[SN]{the}{a} theory 
% of 
behind 
sampling 
operations 
% theory 
on 
% the 
graphs, which leads us  
to the design of critically-sampled 
wavelet-filterbanks 
on graphs. We 
% introduce 
% present
describe
a {\em downsample then upsample} ($DU$) operation 
on graphs in which a set of nodes 
in the graph  are first downsampled (removed) and then 
upsampled (replaced) by inserting zeros.  
This work stems from our 
recent results in~\cite{ICASSP11Sunil}, where we showed 
that downsampling 
% in datasets, based
for graph-signals defined
on {\em $k$-regular bipartite graphs}
% where 
% downsampling  
% recovery of data information 
% loss in the information
% information loss 
% loss of information in the data 
% after these sampling operations 
% on the graph
% are
is governed by a Nyquist-like 
theorem.
%  for graphs.  
% Our key observation 
In this paper, we extend the results 
presented in 
% is that the regularity condition of 
% the degrees of the nodes in 
\cite{ICASSP11Sunil} to all undirected 
bipartite graphs and show that in these graphs, 
% can be dropped 
% and  
% more generally 
% for any undirected
% an important class of graphs 
% known as 
% bipartite graph (i.e. $2$-colorable 
% graph), 
the $DU$ operations
% $G = (\Sc_1,\Sc_2,\textit E)$, 
% the {\em downsampling then upsampling }$DU$ 
% operation 
% We proved  in k-RBG graphs: 
lead to a spectral decomposition 
of the 
% output 
graph-signal 
where
% with  
spectral coefficients are
% being  
reproduced at mirror 
graph-frequencies
% eigenvalues 
around a central frequency.
% eigenvalue.
This is a phenomenon 
% is 
we term
% ed 
as 
{\em spectrum folding in graphs} 
as it is analogous to the frequency-folding 
or ``aliasing'' effect
\change[SN]{in regular signal processing domain}{for regular one-dimensional signals}. 
\change[SN]{We utilize this property to first propose 
ideal  anti-aliasing low-pass filters 
on bipartite graphs 
% We further
and
% We 
show that these anti-aliasing filters, 
when applied 
to a bipartite-graph formulation of regular images, 
are similar in 
form to the well-known regular anti-aliasing 
filters for images. 
% However these filters are global in nature.  
For `real world' applications
we propose
% further exploit these results to design 
critically sampled 
% {\em aliasing cancellation} 
filterbanks for bipartite graphs 
% that
% can be computed locally at each node
% . 
}{We utilize this property to 
propose two-channel filterbanks on bipartite graphs
which are critically sampled
}
and
% we 
provide
% We find 
{\em necessary 
and sufficient} conditions for aliasing cancellation, 
perfect-reconstruction 
and orthogonality in these filterbanks.
%  and 
As a practical solution we 
propose a {\em graph-quadrature mirror 
filterbank} (referred to as graph-QMF) design 
for bipartite graphs which has
% which satisfies 
% the conditions of 
all the above mentioned 
properties. However,
the exact realizations of 
the graph-QMF filters 
% are 
% not 
do not have 
well-localized support on the graph
% support
% operations 
% at each node 
and therefore 
we 
% consider
implement 
polynomial approximations 
of these filters 
which 
% have localized 
are locally supported around each node
% supported on a small local neighborhood at each node but 
% incur 
(at the 
cost of 
small reconstruction 
error and loss of orthogonality). 
% in the spatial domain. 
For arbitrary graphs, we formulate a
bipartite subgraph decomposition problem 
well known to the graph-theory community. 
% The minimum such decomposition a.k.a. {\em biparticity}
% of a graph is linked to {\em chromaticity} of the original graph 
% and is NP-hard in general. 
The decomposition provides us an edge-disjoint collection of $K$
bipartite subgraphs, \change[SN]{ and whose union is the original graph}{each with the same vertex set $\Vc$ and 
whose union is the 
% provide an exact approximation of the 
original graph.} 
% \footnote{A {\em $k$-partite} graph is defined as a graph % $G = (\bigcup_{l=1}^k \Sc_k, {\it E})$, 
% whose vertices can be partitioned (colored) into $k$ disjoint sets (colors) so that no two 
% vertices within the same set (color) are adjacent. The term {\em chromatic number} 
% $\chi(G)$ of a graph refers to smallest such $k$.}. 
% $G$. 
Each of these subgraphs
is then used as a separate ``dimension'' 
to filter and downsample
% . This leads 
leading to a 
\change{multi}{$K$}-dimensional separable 
wavelet filterbank design. To the best of our knowledge 
no such invertible and critically sampled two-channel 
filter-bank designs have been proposed 
for arbitrary graphs before.
% so far.  
The outline for the rest of the paper is as follows: 
we describe the basic framework to understand graph-based transforms  
in  Section~\ref{sec:prelim}. In this section we also describe 
and evaluate some of 
the existing work on 
wavelet-like transforms on graph. 
% and their draw-backs. 
In Section~\ref{sec:system_description}
we 
propose our solution 
% for two-channel critically 
% sampled filterbanks on graph and provide necessary and sufficient 
% conditions for them to be 
% critically sampled, orthogonal and provide 
% perfect reconstruction.  
% In this section, we explain the spectral-folding phenomenon on bipartite graphs 
% which lead to our proposed graph-QMF design. 
and in Section \ref{sec:experiments} we demonstrate the utility of proposed 
filterbanks by conducting some experiments. 
% in 
%  and 
Finally, in Section \ref{sec:conclusions},  we conclude and describe our future work.
% of graphs, graph signals and graph based transforms which 
% creates the framework that we use. 
\section{Preliminaries}
\label{sec:prelim}
% \subsection{Notations}
We use the common convention of 
representing matrices and vectors with 
bold letters,
sets with calligraphic capital letters and scalars with 
normal letters. 
A graph can be denoted as $G = (\Vc,\textit E)$  
with vertices (or nodes) in set $\Vc$ and 
links (or edges) as tuples $(i,j)$ in $\textit E$. 
We only consider 
undirected graphs without self-loops in our work. 
The size of the graph $N = |\Vc|$ 
is the number of nodes and 
geodesic 
% (shortest hop) 
distance metric is given as $d(v,m)$. The
$j$-hop neighborhood  
$\Nc_{j,n} = \{v \in \Vc : d(v,n) \leq j\}$ 
of node $n$ is the set of 
all nodes which are at most 
$j$-\change[SN]{hopping}{hop} distance away from 
node $n$. Algebraically, 
a graph can be represented 
with 
% graph-invariant matrices 
% such as 
the node-node adjacency matrix
$\Am$ such that  
% the graph. 
the element $A(i,j)$ is the weight 
of the edge between node $i$ and $j$ ($0$ 
if no edge). The value $d_i$ is the 
degree of node $i$, which 
% represent 
is
the sum of 
weights of all edges connected to 
node $i$, and  
$\Dm = diag(\{d_{i}\})$ denotes the 
diagonal degree matrix whose $i^{th}$ 
diagonal entry is $d_i$.
%  as 
The Laplacian matrix of the graph 
is defined as $\Lm =\Dm - \Am$ and has a 
normalized form 
% of Laplcian matrix is given 
% by 
$\Lcb =\Id - \Dm^{-1/2}\Am\Dm^{-1/2}$, 
where $\Id$ 
is the identity matrix. We 
% further 
denote $<\fv_1, \fv_2 >$ 
as the inner-product between vectors $\fv_1$ and $\fv_2$.
% although other distance metrics are permissible in our frame-work
% Additional notations required in explaining related work are: 
% let
\subsection{Spatial Representation of Graph Signals}
A graph signal is a real-valued scalar 
function $f:\Vc \to \RR$ defined on 
graph $G=(\Vc,{\it E})$ such that 
$f(v)$ is the sample value of function at vertex $v \in \Vc$.\footnote{
The extension 
to complex or vector sample values $f(v)$ 
is possible but is not 
considered in this work.} 
On a finite graph, the graph-signal 
can be viewed as a sequence or 
% It can also be viewed 
a vector $\fv = [f(0), f(1),...,f(N)]^t$\change[SN]{. 
% However 
Note that}{, where}
the order of arrangement  
of the samples in the vector is arbitrary and 
neighborhood (or nearness) information is provided
separately
by the adjacency matrix $\Am$. 
% can only be infered from .  
% Usually the samples $f(v)$ are scalar values although 
% extension to vectors is also possible. 
% The 
Graph-signals can,
for example, be a set of  
measured values 
by sensor network nodes~\cite{Ramchandran'06} or traffic 
measurement samples on the edges of 
an Internet graph~\cite{Crovella'03} or 
% some 
information 
about the actors in a social network. 
\change[SN]{
A graph based transform is a 
linear operator $\Tm:\RR^N \to \RR^M$ 
applied to $N$-D graph-signal space, 
such that each 
operator output is a linear combination of  
value of the graph-signal $f(n)$ at a node $n$ and the values of samples $f(m)$ on 
nearby nodes $m \in \Nc_{j,n}$.}{Further, a graph based transform is defined as
a linear transform $\Tm:\RR^N \to \RR^M$ 
applied to the $N$-node graph-signal space, 
such that the operation at each 
node $n$ is a linear combination of the
value of the graph-signal $f(n)$ at the node $n$ and the values $f(m)$ on 
nearby nodes $m \in \Nc_{j,n}$,  
}
% its $j$-hop neighborhood, 
i.e.,
\begin{equation}
y(n) = T(n,n) f(n) + \sum_{m \in \Nc_{j,n}} T(n,m) f(m)
 \label{eq:local_linear_tx}
\end{equation}
\add[SN]{In analogy to the $1$-D regular case, we would sometimes refer to graph-transforms as graph-filters and}
the elements $T(n,m)$ for $m = 1,2,...N$ as the filter coefficients at the $n^{th}$ \change{filter}{node}. \remove[SN]{and depend
% Thus the graph transforms are 
% different from regular 
% transforms as they are 
% based 
on the connectivity 
information encoded in 
the edge weights. }
% and are 
% said to 
% be 
% (in other words 
\change[SN]{\protect The transform in (\ref{eq:local_linear_tx})}{A graph transform} is said 
to be  strictly $j$-hop
localized in
the \change[SN]{spatially}{spatial} 
domain of the graph 
\change[SN]{since the 
transform}{if the filter}  coefficients 
$T(n,m)$ are zero 
beyond the $j$-hop 
neighborhood of \add[SN]{each} node $n$. 
\change[SN]{However}{Note that} spatial localization 
can also be applied in a weaker 
sense
% , so that the
in which 
%  if bulk of the mass 
% of $n^{th}$ filter $\Tm(n,:)$ is 
% concentrated in a small $k$-hop 
% neighborhood around node $n$
% and the
% in strict sense 
% if $k$ in (\ref{eq:local_linear_tx}) is small 
% compared to the diameter of the graph or in a
% weak sense if 
\change[SN]{transform}{filter} coefficients 
$T(n,m)$ 
% are not exactly zero but 
decay sharply 
in magnitude
% with increasing distance
% $dist(n,m)$  
beyond $j$-hop neighborhood of node $n$. 
\subsection{Spectral Representation of Graph Signals}

The Laplacians $\Lm$ and $\Lcb$ are both symmetric 
positive semidefinite matrices and therefore\change[SN]{by}{, from the }spectral 
projection theorem, there 
exists a \add[SN]{real} unitary matrix 
$\Um$ which diagonalizes 
$\Lcb$, such that 
$\Um^t\Lcb\Um = \Lambda = diag\{\lambda_i\}$
is a non-negative diagonal matrix.  
\remove[SN]{The eigenvectors $\uv_1, \uv_2, ...,\uv_N$ 
which are columns of $\Um$ form a basis in $\RR^N$ 
and the corresponding
eigenvalues $\sigma(G)= \{0 \leq \lambda_1 \leq \lambda_2 ... \leq \lambda_N\}$ 
represent $N$ orthogonal eigen-spaces $V_{\lambda_i}$ with 
projection matrices  $\uv_i\uv_i^t$.} This leads to an 
{\em eigenvalue decomposition} of matrix $\Lcb$ given as 
\begin{equation}
\Lcb = \Um \Lambda \Um^t = \sum_{i =1}^N \lambda_i \uv_i\uv_i^t ,
 \label{eq:EVD_def}
\end{equation}
\add[SN]{where the eigenvectors $\uv_1, \uv_2, ...,\uv_N$, 
which are columns of $\Um$ form a basis in $\RR^N$ 
and the corresponding
eigenvalues $\sigma(G)= \{0 \leq \lambda_1 \leq \lambda_2 ... \leq \lambda_N \}$ 
represent $N$ orthogonal eigen-spaces $V_{\lambda_i}$ with 
projection matrices  $\uv_i\uv_i^t$.}
Thus, every graph-signal $\fv \in \RR^N$ can 
be decomposed into a linear combination of eigenvectors
$\uv_{i}$ given as $ \fv = \sum_{n = 1}^N \bar{f}(n)\uv_n$.
% \begin{equation}
% 
%  \label{eq:spectral_decompx}
% \end{equation}
It has been shown in~\cite{fchung,luxberg'07} that the 
eigenvectors of Laplacian matrix 
provide a harmonic analysis of 
graph signals which 
gives a
Fourier-like interpretation. \change[SN]{Thus t}{T}he eigenvectors act 
as the {\em natural vibration modes} 
of the graph, and the 
corresponding eigenvalues 
as the associated {\em graph-frequencies}. The {\em spectrum} \add[SN]{$\sigma(G)$} of a graph is 
defined as the set of eigen-values of its normalized Laplacian matrix \add[SN]{\protect and it is always a subset of closed set $[0,2]$ for any graph $G$. 
Any eigenvector $\uv_{\lambda}$ is considered to be a  low pass eigenvector if the magnitude of the
corresponding eigenvalue $\lambda$ is small, i.e., close to $0$. Similarly, 
an eigenvector is  a high-pass eigenvector if its eigenvalue is large, i.e., close to the highest graph-frequency. }\footnote[1]
{The 
% association 
mapping $\uv_n \to \Vc$ associates 
the real numbers $u_n(i), i = \{1,2,...,N\}$, with the vertices $\Vc$ of $G$. 
The numbers $u_n(i)$ will be positive, negative or zero. The frequency interpretation of eigenvectors can thus 
be understood in terms of number of zero-crossings (pair of nodes with different signs) of eigenvector $\uv_n$ on the graph $G$.
For any finite graph the eigenvectors with large eigenvalues have more zero-crossings (hence high-frequency) 
than eigenvectors with small eigenvalues. These results are related to `nodal domain theorems' and
readers are directed to~\cite{NodalDomainThm} for 
more details.}
\change[SN]{\protect The {\em graph Fourier transform }(GFT),  
denoted as $\bar{\fv}$, 
for any signal
$\fv \in \RR^N$ on graph $G$ 
has been defined in \cite{Hammond'09} as}{\protect The {\em graph Fourier transform }(GFT),  
denoted as $\bar{\fv}$, is defined in \cite{Hammond'09} as the projections of a signal $\fv$ on the graph $G$ 
onto the eigenvectors of $G$, i.e., }
%  \add[SN]{the projection of signal $\fv$ onto the graph eigenvectors, i.e.,}
\begin{equation}
\bar{f}(\lambda) = <\uv_{\lambda}~,~ \fv> = \sum_{i = 1}^N f(i)u_{\lambda}(i).
 \label{eq:GFT}
\end{equation}
\change[SN]{\protect The signals whose GFT coefficients have
significant energy $|\bar{f}(l)|^2$ only at 
graph-frequencies close to zero can then be 
called {\em low-pass} graph-signals. Similarly the {\em high-pass} 
graph-signals are the ones which have their significant GFT coefficients located 
close 
to the highest graph-frequency.}{\protect Note that GFT is an energy preserving transform and 
a signal can be considered low-pass (or high-pass) if the energy $|\bar{f}(\lambda)|^2$ of the GFT coefficients is 
mostly concentrated on the low-pass (or high-pass) eigenvectors.}
% Further 
In case of eigenvalues with multiplicity greater than $1$ (say $\lambda_1 = \lambda_2 = \lambda$) 
the 
% EVD is not uniquely defined 
% as 
eigenvectors $\uv_1,\uv_2$ 
are unique up to a unitary transformation in the eigenspace $V_{\lambda} = V_{\lambda_1} = V_{\lambda_2}$.
% can be arbitrarily
% chosen from the $2$-dimensional eigenspace 
% $V_{\lambda_1} = V_{\lambda_2} $. To eliminate 
% this arbitrary choice its better to take 
In this case we can choose 
$\lambda_1\uv_1\uv_1^t + \lambda_2\uv_2\uv_2^t  = \lambda \Pm_{\lambda} $ 
where $\Pm_{\lambda}$ is the projection matrix for eigenspace $V_{\lambda}$. 
Note that for all symmetric matrices, the dimension of eigenspace 
$V_{\lambda}$ (geometric multiplicity)
is equal to the multiplicity of eigenvalue $\lambda$ (algebraic multiplicity) and  
the spectral decomposition in (\ref{eq:EVD_def}) can be
written as 
\begin{equation}
\Lcb = \sum_{\lambda \in \sigma(G)}\lambda \sum_{\lambda_i = \lambda} \uv_i\uv_i^t = \sum_{\lambda \in \sigma(G)}\lambda \Pm_{\lambda}.
 \label{eq:ESD_def}
\end{equation}
\add[SN]{\protect The eigenspace projection matrices are idempotent and $\Pm_{\lambda}$ and $\Pm_{\gamma}$ are orthogonal if $\lambda$ 
and $\gamma$ are distinct eigenvalues of the Laplacian matrix, i.e., 
\begin{equation}
\Pm_{\lambda}\Pm_{\gamma} = \delta(\lambda - \gamma)\Pm_{\lambda},
\label{eq:eigenspace_prop1}
\end{equation}}
where $\delta(\lambda)$ is the Kronecker delta function.
% Further, the graph-signal space in $\RR^N$ is the direct sum of eigenspaces $V_{\lambda}$
% for $\lambda \in \sigma(G)$ and the 
% eigen of can be decomposed into 
% its projection on the eigenspaces $V_{\lambda}$ and
% }
\subsection{Downsampling in Graphs}
% 
% , defined analogous to 
% regular wavelet transforms,  is  
We define the downsampling operation $\betav_H$ on 
the graph $G = (\Vc,\textit E)$ 
% can be 
% defined 
as choosing a subset \change[SN]{$\Sc \subset \Vc$}{$H \subset \Vc$} such that  
% by an integer factor $M$ 
all samples of the graph signal $\fv$, corresponding to indices not in \change[SN]{$\Sc$}{$H$}, are 
discarded. A subsequent upsampling operation 
projects the downsampled signal back to original $\RR^N$ space by
inserting zeros in place of discarded samples in \change[SN]{$\Sc^c$}{$H^c = L$}. 
Given such a set \change[SN]{$\Sc$}{$H$} we define a 
% downsampling 
% this operation to general graphs by defining 
%We define 
{\it downsampling function} 
\change[SN]{$\betav_{\Sc} \in \{-1,+1\}$}{$\betav_{H} \in \{-1,+1\}$} 
% 
% 
% for $\Sc \subset \Vc$  
given as 

\begin{equation}
\displaystyle \beta_{H}(n) =
\left\{
 \begin{array}{ll}
1 & \mbox{if } n \in H\\
-1 & \mbox{if } n \notin H \\
\end{array}
\right.
\label{eq:beta_gen}
\end{equation}

and a diagonal {\em downsampling matrix} \add[SN]{$\Jm_{\beta_H} = diag\{\beta_H(n)\}$}.
The overall  
`downsample then upsample'($DU$) operation  
can then be algebraically represented 
 as 

\begin{equation}
f_{du}(n) = \frac{1}{2}(1 + \beta_{H}(n))f(n)
 \label{eq:one_dim_downsample1}
\end{equation}

and in matrix form as \change[SN]{$\fv_{du} = 1/2(\Id + \Jm_{\beta_{\Sc}})\fv$}{
\protect 
\begin{equation}
\fv_{du} = \frac{1}{2}(\Id + \Jm_{\beta_{H}})\fv 
\label{eq:one_dim_downsample}
\end{equation}}
%  in Figure \ref{fig:graph_al}
%%%%%%%%%%%%%%%%%%%%%%%%%%%%%%%%%%%%%%%%%%%%%%%%%%%%%%%%%%%%%%%%%%%%%%%%%
%%%%%%%%%%%%%%%%%%%%%%%%%%%%%%%%%%%%%%%%%%%%%%%%%%%%%%%%%%%%%%%%%%%%%%%%%%
%SN: Figure Removed
% \begin{figure}[htb]
% \centering
% \includegraphics[width=3in]{dwnsmpl_graph_al.pdf}
% \caption{\footnotesize Downsampling in graphs: algebraic form}
% \label{fig:graph_al}
% \end{figure}
%%%%%%%%%%%%%%%%%%%%%%%%%%%%%%%%%%%%%%%%%%%%%%%%%%%%%%%%%%%%%%%%%%%%%%%%
% \begin{equation}
% \fv_{du} = \frac{1}{2}(\Id + \Jm_{\beta_{\Sc}})\fv
%  \label{eq:graph_mat}
% \end{equation}
%  where 
Note that \add[SN]{$\Jm_{\beta_H}$} is a symmetric matrix such that \add[SN]{$\Jm_{\beta_H}^2 = \Id$} (identity matrix).
\change[SN]
{\protect Since both input and output signals ($\fv$ and $\fv_{du}$) are in $\RR^N$,
% space
% , 
they both have 
a GFT decomposition according to (\ref{eq:GFT}), denoted here by $\bar \fv$ 
and $\bar \fv_{du}$ respectively}
{\protect Since the graph-signal after $DU$ operation also belongs to $\RR^N$,
% space
% , 
it too has 
a GFT decomposition $\bar \fv_{du}$ according to (\ref{eq:GFT})}.
\change[SN]{ A relationship can thus be drawn between the 
output and input spectral coefficients as}{The relationship between the GFTs of 
$\fv$ and $\fv_{du}$ is given as:} 
\add[SN]{ \protect
\begin{equation}
\bar f_{du}(l) = <\uv_l~,~\fv_{du}> = \frac{1}{2}( <\uv_l~,~\fv> + <\uv_l~,~\Jm_{\beta_H}\fv> ) 
 \label{eq:downsampling_spectral_temp}
\end{equation}
}
% = \frac{1}{2}(\bar f(l) + <\Jm_{\beta}\uv_l~,~\fv> ) = \frac{1}{2}(\bar f(l) + \bar f^d(l))
The inner-product \add[SN]{$<\uv_l~,~\Jm_{\beta_H}\fv>$ can also be written as $<\Jm_{\beta_H}\uv_l~,~\fv>$, which}  
represents the projection of input signal $\fv$ onto a {\em deformed} 
eigenvector \add[SN]{$\Jm_{\beta_H}\uv_l$}. We \add[SN]{define this projection} as a {\em deformed} spectral coefficient \add[SN]{$\fv^d(l)$}
and  (\ref{eq:downsampling_spectral_temp}) can be written as: 
%(\ref{eq:downsampling_spectral_temp}) write it as : 
\add[SN]{ \protect
\begin{equation}
\bar f_{du}(l)= \frac{1}{2}(\bar f(l) + <\Jm_{\beta_H}\uv_l~,~\fv> ) = \frac{1}{2}(\bar f(l) + \bar f^d(l))
 \label{eq:downsampling_spectral}
\end{equation}
}
\change[SN]{The 
properties and design of these deformed basis functions form a crucial 
part of our framework.}{ \protect In case of bipartite graphs, the spectrum of the graph
is symmetric and 
the deformed eigenvectors are also the eigenvectors of the same graph.  
This phenomenon, termed as {\em spectral folding}, forms the basis of our two-channel 
filterbank framework, and will be described in detail in Section \ref{sec:system_description}. }  

\subsection{Two-Channel Filterbanks on Graph}
\label{sec:downsampling}
\begin{figure*}[htb]
\centering
\includegraphics[width=4in]{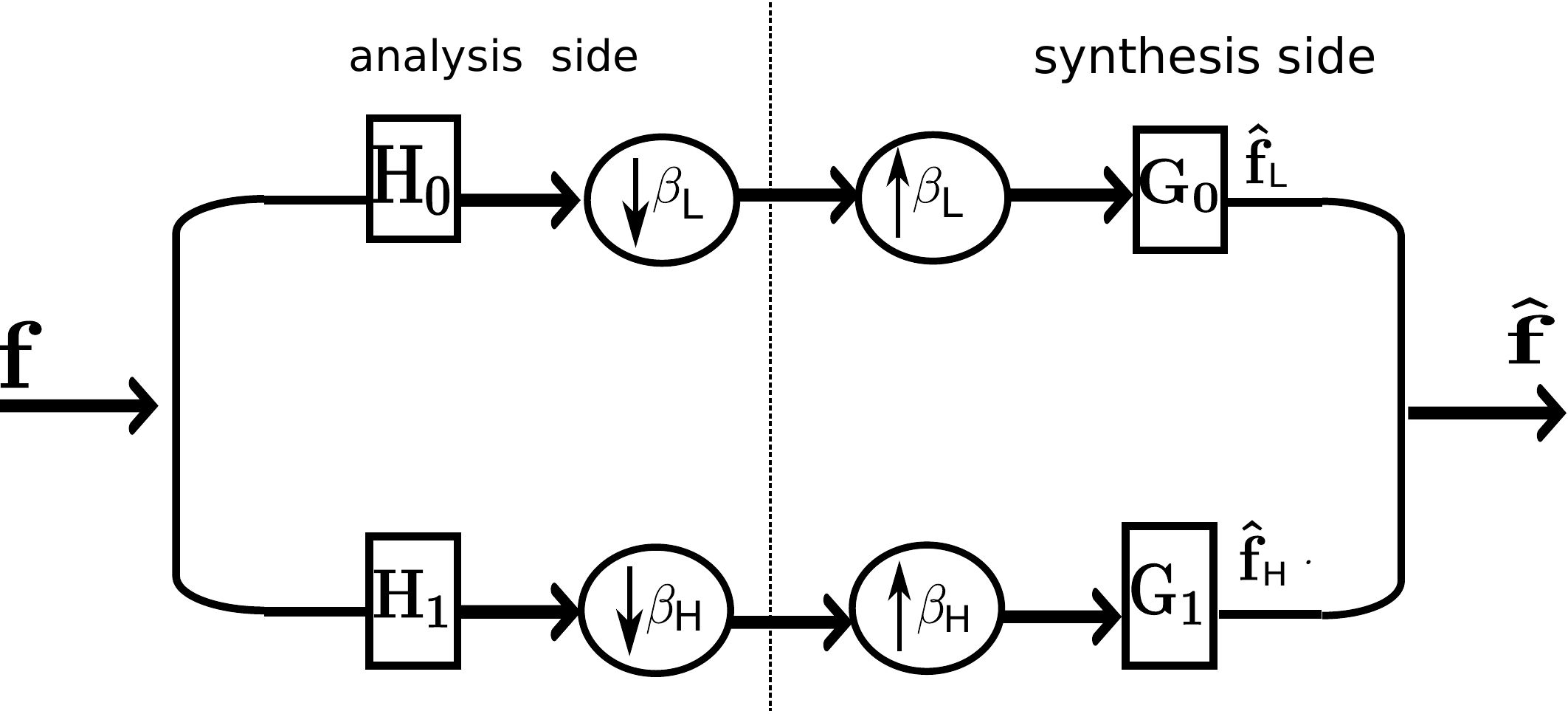}
\caption{\footnotesize Block diagram of a two-channel wavelet filterbank on graph. 
% The  graph transform pair $\{\Hm_{0},\Hm_1\}$ forms the analysis low-pass and analysis high-pass 
% wavelet filters respectively and $\{\Gm_{0},\Gm_1\}$ are corresponding synthesis filters. The downsampling 
% function $\beta$ is such that nodes with $\beta(n) = 1$ store the output of only the low-pass filter and nodes 
% with $\beta(n) = -1$ store the output of only high-pass channel. As a result the transform is critically sampled.
}
\label{fig:graph_filterbank}
\end{figure*}

A 
% critically sampled 
two-channel wavelet filterbank on \add[SN]{a} graph
%  as shown in Figure \ref{fig:graph_filterbank}, 
provides a 
decomposition of any graph-signal into 
a lowpass \change[SN]{(approximation)}{(smooth)} graph-signal
and a highpass (detail) graph-signal component.
%SN: lowpass and highpass graph signals have already been defined  above
\change[SN]{ The transforms $\Hm_i,i=\{0,1\}$ of the two channels contain 
spatially localized filters}{ \protect The two channels of the 
filterbanks
% of the
% critically sampled 
% two-channel wavelet filterbank 
are 
characterized by the graph-filters 
$\{\Hm_i,\Gm_i\}_{i \in \{0,1\}}$ and the downsampling operations $\betav_{H}$ and 
$\betav_{L}$ 
% and $\beta_{\Sc_2}(n)$ 
as shown in Figure \ref{fig:graph_filterbank}.
}
% , provides a
% decomposition of graph-signal $\fv$ into 
% a lowpass (approximation) graph-signal $\fv_{low}$
% and a highpass (details) graph-signal component $\fv_{high}$. 
% The transforms $\Hm_i,\Gm_i$ of the two channels 
% are graph transforms 
% containing  
% spatially localized filters.} and are graph-frequency selective
% designed 
% such that 
The
% This implies that
% lowpass 
transform $\Hm_0$ acts as \add[SN]{a} lowpass \change[SN]{transform}{filter}, i.e., it 
% which 
transfers \add[SN]{the contributions of the low-pass} graph-frequencies 
\remove[SN]{which are close to zero}which are below some cut-off and attenuates significantly
the graph-frequencies which are above the cut-off. The highpass transform
% a lowpass component of the graph-signal and 
% mostly the components of any graph-signal corresponding 
% to graph-frequencies close to zero (i.e. its a lowpass transform), while 
% and 
$\Hm_1$ does the opposite of a low-pass transform, i.e, it
attenuates significantly, the graph-frequencies below some cut-off frequency.
% those frequencies and 
% transfers
% graph-frequencies, which are close 
% to the highest graph-frequency, to pass through.
% it transfers a highpass component of the graph-signal.
% components corresponding to graph-frequencies close to the 
% highest graph-frequency.
% only 
% the 
% lowpass component of any graph-signal while 
% % and a highpass 
% the transform $\Hm_1$ transfers only 
% for 
% the
% highpass component. 
\change[SN]{ The transformation in 
each channel is followed by a
% the
% downsampling then upsampling ($DU$) 
$DU$ operation
using downsampling functions $\beta_{\Sc_1}$ 
and $\beta_{\Sc_2}$ and corresponding downsampling matrices 
$\Jm_{\beta_{\Sc_1}}$ and $\Jm_{\beta_{\Sc_2}}$ 
for lowpass and the highpass channel respectively.}
{The filtering operations in each channel are followed by 
downsampling operations $\betav_{H}$ and $\betav_{L}$,}
\change[SN]{
This}{which 
means that the nodes with membership in
the set} \change[SN]{$\Sc_1$}{$H$} store the 
output of highpass channel while 
the nodes in the 
% compliment 
set \change[SN]{$\Sc_2$}{$L$} 
store
the output of lowpass channel. 
\change[SN]{and 
for critically sampled output ($|\Sc_1| + |\Sc_2| = N$)}{For critically sampled output we have: $|H| + |L| = N$.}\footnote[2]{ Note that in the regular signal domain the 
two most common patterns of critically sampled output 
are i) \change[SN]{$\Sc_1 = \Sc_2 = \{0,2,4,...\}$}{$H = L = \{0,2,4,...\}$}, 
where even set of nodes store the output of both channels and ii) \change[SN]{$\Sc_1 = \{0,2,4,...\}$ and 
 $\Sc_1 = \{1,3,5,...\}$}{$L = \{0,2,4,...\}$ and 
 $H = \{1,3,5,...\}$} , 
where each node stores the output of only one of the channel.}
% , 
% as shown in 
%   
% % What also 
% % becomes apparent from this definition is the notion of 
% % low-pass and high-pass filters. 
% In this context, the {\em low-pass graph
% filters} are defined as 
% transforms whose response is bandlimited 
% in the region of the lowest graph-frequency bands whereas
% the {\em high-pass graph filters} are  defined as transforms 
% whose response is bandlimited in the highest 
% graph-frequency bands. 
% Now with the definition of wavelet filters and downsampling/upsampling 
% operations on graph in our hands, .
% 
% 
% We choose a downsampling function $\beta = \beta_{\Sc}$ for one of the channels which 
% % according to our definition 
% assigns 
% $1$ to all the nodes in the set $\Sc$ and $-1$ to the 
% remaining nodes in the set $\Sc^c$. 
% The downsampling function defined on the other channel is $-\beta(n) = \beta_{\Sc^c}(n)$ which is 
% the sign-reversed 
% form of the original downsampling function $\beta_{\Sc}(n)$. 
% Note that the downsampling function for the 
% highpass channel is $-\beta(n) = \beta_{\Sc^c}(n)$ which 
% in both channels have different signs. 
% thus making the overall transform critically sampled. 
Using (\ref{eq:one_dim_downsample}), it is easy to see from Figure \ref{fig:graph_filterbank} that the output signals 
in the lowpass and highpass channels, after reconstruction 
are given as 
% \change[SN]{$1/2(\Id + \Jm_{\beta_{\Sc_1}})\Hm_0\fv$ 
% and $1/2(\Id + \Jm_{\beta_{\Sc_2}})\Hm_1\fv$}{$1/2(\Id + \Jm_{\beta_{L}})\Hm_0\fv$ 
% and $1/2(\Id + \Jm_{\beta_{H}})\Hm_1\fv$} and the outputs after applying synthesis 
% filters $\Gm_0$ and $\Gm_1$ respectively are given as 
\begin{eqnarray}
\hat \fv_{L} &= &\frac{1}{2}\Gm_0(\Id + \Jm_{\beta_{L}})\Hm_0\fv \nonumber \\
\hat \fv_{H} &=& \frac{1}{2}\Gm_1(\Id + \Jm_{\beta_{H}})\Hm_1\fv,
\label{eq:channel_outp}
\end{eqnarray}
respectively.
The overall output $\hat \fv$ of the filterbank is the sum of outputs of the two channels, i.e., 
$\hat \fv= \hat \fv_{L} + \hat \fv_{H} = \Tm  \fv$, 
where $\Tm$ is the overall transfer function of the filterbank given as:
% and summing the outputs is given as:
% $\Tm_{a}$ on the analysis side  is an $N \times N$ square matrix 
% given as:
% \change[SN]{\protect
% \begin{eqnarray}
%  \displaystyle \Tm & = &  \frac{1}{2}\Gm_0(\Id + \Jm_{\beta_{\Sc_1}})\Hm_0 + \frac{1}{2}\Gm_1(\Id + \Jm_{\beta_{\Sc_2}})\Hm_1  \nonumber \\
% & = &  \underbrace{\frac{1}{2}(\Gm_0\Hm_0 + \Gm_1\Hm_1)}_{A} +  \underbrace{\frac{1}{2}(\Gm_0\Jm_{\beta_{\Sc_1}}\Hm_0 + \Gm_1\Jm_{\beta_{\Sc_2}}\Hm_1)}_{B}
% % \label{eq:overall_tx}
% \end{eqnarray}
% }{\protect
\begin{eqnarray}
 \displaystyle \Tm & = &  \frac{1}{2}\Gm_0(\Id + \Jm_{\beta_{L}})\Hm_0 + \frac{1}{2}\Gm_1(\Id + \Jm_{\beta_{H}})\Hm_1  \nonumber \\
& = &  \underbrace{\frac{1}{2}(\Gm_0\Hm_0 + \Gm_1\Hm_1)}_{\Tm_{eq}} +  \underbrace{\frac{1}{2}(\Gm_0\Jm_{\beta_{L}}\Hm_0 + \Gm_1\Jm_{\beta_{H}}\Hm_1)}_{\Tm_{alias}}, 
\label{eq:overall_tx}
\end{eqnarray}
% }
\change[SN]{\protect In (\ref{eq:overall_tx})}{where} 
\remove[SN]{the expression denoted as} $\Tm_{eq}$ is the transfer function of the filterbank 
without the $DU$ operation and \remove[SN]{
the expression} $\Tm_{alias}$ is another transform which arises primarily due to the \change[SN]{$DU$ operations}{downsampling in the two channels}. 
\change[SN]{For perfect reconstruction 
$\Tm$ should be equal to identity which means \remove[SN]{term} $A$ should be a scalar multiple of identity and \remove[SN]{term} $B$ should cancel out.}
{ For perfect reconstruction $\Tm$ should be equal to identity which can be ensured by
requiring $\Tm_{eq}$ to be a scalar multiple of identity and $\Tm_{alias}= \zerov$.}
% :
% 
% \begin{equation}
%  \underbrace{(\Gm_0\Hm_0 + \Gm_1\Hm_1)}_{response} +  \underbrace{(\Gm_0\Jm_{\beta_{\Sc_1}}\Hm_0 + \Gm_1\Jm_{\beta_{\Sc_2}}\Hm_1)}_{deformation} = \Id
% \label{eq:overall_tx}
% \end{equation}
% T
% the deformation term in (\ref{eq:overall_tx}), which arises primarily due to the $DU$ operations should cancel out.
%  and the response term in 
% (\ref{eq:overall_tx}) should be identity.
% 
% and it is desirable to choose downsampling functions 
% $\beta_{\Sc_1}$ and $\beta_{\Sc_2}$ in such a way that the deformation term cancels out. 
{\em Thus the two-channel filterbank on a graph provides distortion-free  
perfect reconstruction if}
% \change[SN]{\protect
% \begin{equation}
% \begin{array}{l}
% \Gm_0\Jm_{\beta_{\Sc_1}}\Hm_0 + \Gm_1\Jm_{\beta_{\Sc_2}}\Hm_1  = \zerov \\ 
%  \Gm_0\Hm_0 + \Gm_1\Hm_1 = c\Id \\
% \end{array}
% \label{eq:general_PR}
% \end{equation}
% }{\protect
\begin{eqnarray}
\Gm_0\Jm_{\beta_{L}}\Hm_0 + \Gm_1\Jm_{\beta_{H}}\Hm_1  & = & \zerov \nonumber \\ 
 \Gm_0\Hm_0 + \Gm_1\Hm_1 & = & c\Id 
\label{eq:general_PR}
\end{eqnarray}
% }
% 
% Looking at (\ref{eq:overall_tx}) we find that $(\Gm_0\Jm_{\beta_{\Sc_1}}\Hm_0 + \Gm_1\Jm_{\beta_{\Sc_2}}\Hm_1)$ is a deformation in the 
% arises due to the $DU$ operation and 

% \begin{equation}
%  \begin{array}{l}
% \Gm_0 = \frac{1}{2}(\Id + \Jm_{\beta})\Tm_{s}^t \\ \nonumber
% \Gm_1 = \frac{1}{2}(\Id - \Jm_{\beta})\Tm_{s}^t
% \end{array}
% \label{eq:spectral_syn}
% \end{equation}
In order to \change[SN]{realize these}{design perfect reconstruction} filterbanks we need 
% the knowledge of
to determine
a) how to design filtering operations $\Hm_i,\Gm_i, i=\{0,1\}$, 
% i.e. graph transforms which are 
% localized in both spatial and spectral domain of the graph and 
and
b) \remove[SN]{how to choose} the downsampling functions \change[SN]{$\beta_{\Sc_1}$ and $\beta_{\Sc_2}$}{$\beta_{L}$ and $\beta_{H}$}. 
In Section \ref{sec:system_description}, \change[SN]{\protect define the desirable conditions for 
designing these transforms and downsampling functions. 
}{\protect we show that the spectral folding phenomenon 
in bipartite graphs leads to an aliasing interpretation of (\ref{eq:general_PR}) and we 
design filterbanks which cancel aliasing and lead to perfect reconstruction of any graph-signal. }
% which 
% together with the designed transforms, 
% satisfies (\ref{eq:overall_tx}).
% are {\em aliased and shifted}
% copy of 
% the original spectrum $\bar \fv$ and c)
% %  how to design  
% % coefficients $\bar \fv$ and can be recovered by ``anti-aliasing'' filters and c) 
% what are the 
% {\em perfect reconstruction}(PR) conditions for which reconstructed signal $\hat \fv = \fv$. 
Before 
explaining our approach, 
% 
% A two-channel wavelet filterbank 
% () on graph consists 
% 
we \add[SN]{briefly} analyze and evaluate some of the existing 
% research 
% on 
graph based 
transforms\add[SN]{, by representing them using the framework we just introduced}.
%  and show that they can also be represented in terms of our framework. 

\subsection{Existing Designs}
\label{sec:lit_review}
% In this subsection, we analyze 
% some of the 
Existing designs of wavelet-like filterbanks 
on the graph 
% based on the framework we have 
% discussed so far in this section. 
% These 
% transforms 
can be divided into two types, namely,
% a) those 
% which are based on 
spatial and spectral designs. In order to understand these designs we \change[SN]{define}{introduce}
some additional notation\change[SN]{s given as: w}{.W}e define  
$\partial \Nc_{h,k}$ to be an $h$-hop neighborhood ring around node $k$ (i.e., the set of all 
nodes which are exactly $h$ hops away from node $k$), a $j$-hop adjacency matrix $\Am_j$ s.t. 
$\displaystyle \Am_{j}(n,m) = 1 ~\mbox{only if } m \in \Nc_{j,n}$, 
a $j$-hop \add[SN]{diagonal} degree matrix  with \change[SN]{$\Dm_{j}= diag\{d_{j,k}\}$ s.t. $d_{j,k} = |\Nc_{j,k}|$ }{$D_j(k,k)=|\Nc_{j,k}|$ s.t. $d_{j,k} = |\Nc_{j,k}|$} and
%  $\Lm_{j}$ as 
a $j$-hop uniform Laplacian 
matrix 
% given as 
$\Lm_j = \Dm_{j} - \Am_{j}$. 
Similarly \add[SN]{we} define a ring adjacency matrix $\partial \Am_h$ such 
that $\displaystyle \partial \Am_{j}(n,m) = 1 ~\mbox{only if } m \in \partial \Nc_{j,n}$ and 
corresponding ring degree matrix 
$\partial \Dm_h = diag\{\partial d_{j,k}\}$ s.t. $\partial d_{j,k} = |\partial \Nc_{h,k}|$.
% representation and 
% b) those which 
% are based on spectral representation. 
% \
\subsubsection{Spatial Designs}  
Wang and Ramchandran~\cite{Ramchandran'06} 
%who 
proposed spatially localized 
graph transforms for sensor network graphs with binary links\add[SN]{, (i.e. links which have weight either $0$ or $1$)}. 
 The transforms 
proposed in~\cite{Ramchandran'06} 
% are of two types: those which 
% 
% 
% These transforms come in two flavors and 
either compute 
% either 
a weighted 
average given as
\begin{equation}
\displaystyle y(n) = (1-a+\frac{a}{d_{j,k}+1})x(n) + \sum_{m \in \Nc_{j,n}} \frac{a}{d_{j,k}+1}x(m),
% 
% 
% f_{j,k}(n) =
% \left\{
%  \begin{array}{ll}
% 1- a + \frac{a}{d_{j,k}+1} & \mbox{if } n =k\\
% \frac{a}{d_{j,k}+1} & \mbox{if } n \in \Nc{j,k} \\
% 0 & \mbox{if } otherwise \\
% \end{array}
% \right.
 \label{eq:average-filter}
\end{equation}
or 
% those which compute 
a weighted difference given as
\begin{equation}
\displaystyle y(n) = (1+ b - \frac{b}{d_{j,k}+1})x(n) - \sum_{m \in \Nc_{j,n}}\frac{b}{d_{j,k}+1}x(m),
% 
% g_{j,k}(n) =
% \left\{
%  \begin{array}{ll}
% 1+ b - \frac{b}{d_{j,k}+1} & \mbox{if } n =k\\
% -\frac{b}{d_{j,k}+1} & \mbox{if } n \in \Nc{j,k} \\
% 0 & \mbox{if } otherwise \\
% \end{array}
% \right.
\label{eq:difference-filter}
\end{equation}
in a  
$j$-hop 
neighborhood around each node in the graph. The corresponding transform matrices can be represented 
% These transforms ~(\ref{eq:average-filter}) and ~(\ref{eq:difference-filter}) can be 
% compared with our definition of a $j$-hop localized graph-transform in ~(\ref{eq:local_linear_tx}). 
for a \change[SN]{fixed 
value of}{given} $j$ 
% and 
as 
\begin{equation}
\begin{array}{l}
\displaystyle \Tm_j = \Id - a (\Id + \Dm_j)^{-1} \Lm_j\\
\displaystyle \Sm_j = \Id + b (\Id + \Dm_j)^{-1} \Lm_j.
\end{array} 
\label{eq:ramchandran_filters}
\end{equation}
% and are invertible 
% for the parameter set $\{0 < a < 1/2,b>0\}$ and 
% The 
% Both type of transforms
% (weighted average or weighted difference)
% are invertible and critically sampled. 
This approach intuitively defines 
a two-channel wavelet filter-bank 
on the graph \remove[SN]{to be}
consisting of 
% In particular we focus on 
two types of linear filters: 
a) {\it approximation filters}\change[SN]{
% those 
which 
compute a weighted average of 
the local neighborhood 
of a given node}{\protect as given in (\ref{eq:average-filter})} and b) 
{\it detail filters}\change[SN]{
% those 
which compute a weighted 
difference of the 
local neighborhood of the node}{ \protect as given in (\ref{eq:difference-filter})}. 
However,
\remove[SN]{ the problem 
in }
these transforms 
% of ~\cite{Ramchandran'06} 
\change[SN]{is that 
they }
are oversampled and
produce output of the size twice that of the input. 
Further none of the transforms can be called a wavelet filter 
since both transforms have a non-zero DC response. 

Crovella and Kolaczyk ~\cite{Crovella'03} designed
wavelet like transforms on graphs 
which are localized in space. 
They defined a collection of functions $\psi_{j,n}: \Vc \to \mathbb{R}$, localized with
respect to a range of scale/location indices $(j, n)$, which 
at a minimum satisfy $\sum_{m \in \Vc}\psi_{j,n}(m) = 0$ 
% \begin{equation}
% \displaystyle 
%  \label{eq:crovella_wavelet_cond1}
% \end{equation}
% or $\psi_{j,k}^{T}\onev = 0$ 
(i.e. a zero DC response). Each function $\psi_{j,n}$ is constant within 
hop rings $\partial \Nc_{h,n}$ and can be written as:
\begin{equation}
\displaystyle y(n) = a_{j,0}x(n) + \sum_{h=1}^j\sum_{m \in \Nc_{h,n}}\frac{a_{j,h}}{\partial d_{j,n}}x(m)
 \label{eq:crovella_filters_al}
\end{equation}
In matrix form the 
$j$-hop wavelet transform $\Tm_j$ can be written as:
\begin{equation}
\Tm_{j} = a_{j,0} \Id + a_{j,1} \partial \Dm_1^{-1} \partial \Am_1 + ... a_{j,j} \partial \Dm_j^{-1} \partial \Am_j
 \label{eq:crovella_filters}
\end{equation}
\change[SN]{For zero DC response condition of these wavelet filters, the constants $a_{j,h}$ 
satisfy $\sum_{h = 0}^{h = j} a_{j,h} = 0$}{Further, the constants $a_{j,h}$ 
satisfy $\sum_{h = 0}^{h = j} a_{j,h} = 0$, which allows the wavelet filters to have zero DC response.}
% \end{equation}
% These coefficients 
\remove[SN]{and can be computed from any continuous wavelet function $\psi(x)$ supported on the interval $[0,1)$ 
% such that 
% $\displaystyle \int_{0}^{1} \psi(x)dx = 0$. 
by taking  $a_{j,h}$ to 
be the average of $\psi(x)$ on the sub-intervals $I_{j,h} = [\frac{h}{j+1}, \frac{h+1}{j+1}]$}. 
% and take $a_h$ to 
% be the average of $\psi(x)$ on interval $I_{j,h}$, i.e 
% \begin{equation}
% a_h = \frac{1}{j+1}\int_{I_{j,h}}\psi(x)d(x)
%  \label{eq:crovella_coeffs}
% \end{equation}
% These graph functions $\psi_{j,k}$ are 
% composed of either shifts or dilations 
% of a single generating function $\psi$.
Though these transforms are local and 
provide a multi-scale summarized view of 
the graph, they do not have approximation filters 
and are not 
invertible in general. 
% In Section ~\ref{sec:related_work} 
% we describe  

Lifting based wavelet transforms for graphs have been 
proposed in~\cite{Wagner'06,Godwin'08,Narang:APSIPA'09, Silverman}
% for graphs 
% in Euclidean Space and 
% for trees in~\cite{Godwin'08}. 
% The {\bf first approach} for critical sampling, is a lifting scheme as
% %the input graph-signal is first decomposed into smaller signals. 
% %Lifting schemes 
% proposed in 
% \cite{Wagner'05, Wagner'06,Godwin'08,Narang:APSIPA'09}, 
% %are an example of this kind of approach 
% which 
and provide
%the second approach for critical 
a natural way of constructing 
local 
two-channel
critically sampled filter-banks on graph-signals. 
% Lifting based transforms are 
% useful for distributed network applications, since
% the inverse lifting transform is also 
% local by construction. 
% A block-diagram 
% of lifting wavelet filter-bank is shown 
% in Figure \ref{fig:fb_bpt_spectral}. 
% \begin{figure}[htb]
% \centering
% \includegraphics[width=4in, height = 1.5in]{block_digram_lifting.pdf}
% \caption[Block diagram for two-channel spectral filter-banks on bipartite graphs]{\footnotesize
% Block diagram for two-channel lifting wavelet filter-banks}
% \label{fig:fb_bpt_spectral}%
% \end{figure}
In this approach the vertex set is 
first partitioned into sets of even and 
odd nodes  
$\Vc = \Oc \cup \Ec$. The odd nodes 
compute their {\em prediction} coefficients using 
their own data and data from their even neighbors
followed by even nodes computing 
their {\em update} coefficients 
using their own data and 
prediction coefficient of their 
neighboring odd nodes. 
The equivalent 
transform in matrix-form 
can be written as:
\begin{equation}
\begin{gathered}
  \Tm^{lift}  = \overbrace {\left[ {\begin{array}{*{20}c}
   {\Id_{\Oc} } & 0  \\
   \Um & {\Dm_{\Ec} }  \\
 \end{array} } \right]}^{update}\overbrace {\left[ {\begin{array}{*{20}c}
   {\Dm_{\Oc} } & { - \Pm}  \\
   0 & {\Id_{\Ec} }  \\
 \end{array} } \right]}^{predict} \hfill \\
%   \Rightarrow \Tm^{lift}  = \left[ {\begin{array}{*{20}c}
%   {\Dm_{\Oc} } & { - \Pm}  \\
%   {\Um\Dm_{\Oc} } & { - \Um\Pm + \Dm_{\Ec} }  \\
% \end{array} } \right]\begin{array}{*{20}c}
%   {\left. { \leftarrow \hat \Tm_{high} } \right\}}  \\
%   {\left. { \leftarrow \hat \Tm_{low} } \right\}}  \\
% \end{array}  \hfill \\  
\end{gathered} 
\label{eq:Lifting_mat}
\end{equation}
where $\Dm_{\Oc}$ and $\Dm_{\Ec}$ are diagonal matrices 
%and $\Id_{\Ec}$ and $\Id_{\Oc}$ are identity matrices of 
of size $|\Oc|$ and $|\Ec|$ respectively.
%The {\it predict} matrix and {\it update}
%matrix are 
%usually 1-hop localized in space. 
%(although general k-hop localized 
%transforms are possible). 
%Therefore it makes sense
%to design lifting 
%based transforms 
%in the spectral domain 
%in the same way as 
%described in previous 
%sections. 
Although the lifting 
scheme can be applied to any 
arbitrary graph, the 
%The implementation 
%of spectral kernel based 
%transforms 
%described 
%above is difficult 
%in the 
%lifting case  
%for two reasons
%a)
%lifting 
design 
% requires 
is equivalent to
simplification
of the graph to 
a bipartite (2-colorable) 
graph\change[SN]{and}{, given that} nodes of the same 
color/parity cannot use 
each 
other's data even 
if they are connected 
by an edge. This results 
in edge losses. 
% Wavelet lifting transforms are guaranteed to be 
% invertible, as long as 
% nodes in the graph 
% are partitioned into two sets 
% (even and odd nodes) and 
% the transform is 
% structured by transforming odd nodes based on 
% only even nodes (and vice versa). These transforms 
% are critically sampled and 
% invertible by construction. However, in 
% lifting scheme same parity nodes do not use each other's data 
% to compute transform hence the links 
% between same parity nodes are essentially 
% lost. 
% most of these approaches, 
% the even/odd splitting and 
% transform design 
% are based on heuristics.
\subsubsection{Spectral Designs}
Maggioni and Coifman \cite{Coifman'06} introduced ''diffusion wavelets'', 
a general theory for wavelet decompositions based 
on compressed representations of 
powers of a diffusion
operator (such as Laplacian). 
Their construction interacts with the underlying
graph or manifold space through 
repeated applications of a diffusion 
operator $\Tm$, such as 
the graph Laplacian $\Lm$.  
% basis functions at coarser 
% resolution level 
% are obtained through a 
% variation of the
% Gram-Schmidt(GSM) 
% orthonormalization scheme 
The localized basis functions at each resolution level are 
orthogonalized and downsampled appropriately to transform sets
of orthonormal basis functions through a 
variation of the
Gram-Schmidt
orthonormalization (GSM) scheme.
Although this local GSM method orthogonalizes the 
basis functions (filters) into  well localized `bump-functions' 
in the spatial domain, it does not provide 
guarantees on the size of the support
of the filters it constructs.  
% to analyze. 
Further 
% However, 
the diffusion wavelets form an over-complete basis\change[SN]{ and 
one has to appropriately choose a 
subset of bases for better approximation}{ and there is no simple way of representing
the corresponding transform $\Tm$}. 

Hammond et al \cite{Hammond'09} defined spectral 
graph wavelet transforms that are determined by 
the choice of a kernel function 
$g: \mathbb{R}^+ \to \mathbb{R}^+$. The kernel 
$g(\lambda)$ is a continuous bandpass 
function in spectral domain with $g(0) = 0$ and 
$\lim_{\lambda \to\infty}g(\lambda)=0$. 
The 
corresponding wavelet operator 
$\Tm_{g} = g(\Lcb) = \Um g(\Lambda)\Um^t$
acts on a graph signal 
$\fv$ by modulating each 
Fourier mode as
\begin{equation}
 \Tm_g\fv = \sum_{k = 1}^N g(\lambda_k)\bar f(k) \uv_k 
\end{equation}
The kernel can be scaled as $g(t\lambda)$ by a 
continuous scalar $t$.  
For spatial localization, 
% Hammond et al \cite{Hammond'09}
the authors 
design \remove[SN]{approximate}
filters by  
approximating the kernels 
$g(\lambda)$ with smooth polynomials 
functions. 
% It is easy to 
% see that 
The approximate 
transform with 
polynomial kernel of degree $k$ 
is given by $\Tm_{\hat g} =
\hat g(L) = \sum\nolimits_{l = 0}^k {a_l L^l }$  
and is exactly $k$-hop localized in space.
By construction the 
spectral wavelet transforms have zero DC response, hence 
in order to stably represent the low frequency 
content of signal $\fv$ a second class of kernel function $h : \RR^+ \to \RR^+$ is introduced 
which acts as a lowpass filter, 
and satisfies $h(0) > 0$ and $\lim_{\lambda \to \infty}h(\lambda)= 0$. Thus a 
multi-channel 
% oversampled 
wavelet transform can be 
% designed Further they show that 
% for a multi-channel transform 
constructed from the choice of a low pass kernel $h(\lambda)$ and 
$J$ band-pass kernels $\{g(t_1\lambda),...,g(t_J\lambda)\}$ 
% spectral 
% kernels  at $J$ scales 
and 
it is been shown that the 
perfect reconstruction  
of the original signal is assured if the quantity 
$G(\lambda)= h(\lambda)^2 +\sum_{k = 1}^J g(t_i\lambda)^2 > 0$
on the spectrum of $\Lcb$ (i.e., at the $N$ eigenvalues of $\Lcb$). 
However, these 
transforms are overcomplete, 
for example, 
a $J$-scale
decomposition 
of graph-signal 
of size 
$N$ \change[SN]{outputs transform coefficients
of size $(J+1)\times N$}{produces $(J+1)N$ transform coefficients}. 
As a result, \remove[SN]{this 
makes}the transform is invertible 
only by the least square 
projection of the output signal onto 
a lower dimension subspace.
%  Representative
% For a given spectral wavelet kernel $g$,
% the wavelet operator $\Tm_{g} = g(\Lcbm)$ acts on a graph signal 
% $f$ by modulating each Fourier mode as
% ˆ
% Tg f ( ) = g(λl )f ( )

To conclude this section, Table~\ref{tab:tab1} presents a summary of existing methods and their properties.
\begin{table}[htb]
\centering
\begin{tabular}{|p{3.5 cm}|p{3 cm}|p{1.5 cm}|p{2 cm}|c|p{2 cm}|}
\hline
% \newline
Method & DC response & Critical Sampling & Perfect Reconstruction & Orthogonality & Requires Graph Simplification \\ 
\hline
% \newline
Wang \& Ramchandran~\cite{Ramchandran'06} & non-zero & No & Yes & No & No \\ 
\hline
% \newline
Crovella \& Kolaczyk~\cite{Crovella'03} & zero  & No & No & No & No \\ 
\hline
% \newline
Lifting Scheme~\cite{Narang:APSIPA'09} & zero for wavelet basis & Yes & Yes & No & Yes \\ 
\hline
% \newline
Diffusion Wavelets~\cite{Coifman'06} & zero for wavelet basis & No & Yes & Yes & No \\ 
\hline
% \newline
Spectral Wavelets~\cite{Hammond'09} & zero for wavelet basis & No & Yes & No & No \\ 
\hline
Proposed graph-QMF filterbanks & zero (when degree-normalized) for wavelet basis & Yes  & Yes & Yes & No \\ 
\hline
\end{tabular}
\caption{\footnotesize Evaluation of existing graph wavelet transforms.}
\label{tab:tab1}
\end{table}%
\footnote{Our proposed solutions can be perfect reconstruction 
and orthogonal without being local or can be local with approximate reconstruction.}
\section{Proposed Solution}
\label{sec:system_description}
% The block 
% diagram of 
Our proposed two-channel critically 
sampled graph wavelet filterbanks are
% same as 
shown in 
Figure~\ref{fig:graph_filterbank}. 
For $DU$ operations we choose a specific downsampling pattern in which 
sets \change[SN]{$\Sc_1$ and $\Sc_2$}{$H$ and $L$} provide a bipartition of the graph nodes
% are mutually disjoint and exhaustive 
\change[SN]{(i.e., $\Sc_1 \cap \Sc_2 = \phi$ and $\Sc_1 \cup \Sc_2 = \Vc$)}{i.e., $H \cap L = \phi$ and $H \cup L = \Vc$)}. 
This implies that downsampling functions 
\change[SN]{
$\beta_{\Sc_1}(n) = - \beta_{\Sc_2}(n) = \beta(n)$}{$\beta_{L}(n) = - \beta_{H}(n) = \beta(n)$} \change[SN]{. Thus}{and the} nodes in \change[SN]{
$\Sc_1$}{$L$} 
store the output of 
\change[SN]{only}{the} lowpass channel 
whereas \add[SN]{the} nodes in the complement 
set \change[SN]{$\Sc_2 = \Sc_1^c = \Vc - \Sc_1$}{$H$} store the output 
of \change[SN]{only}{the} highpass channel\change[SN]{ thus 
making the overall output critically sampled.}{. The overall output after filtering and downsampling operations in both 
channels is critically sampled.}
% The
% two crucial elements in the
% construction of these filterbanks are
% i) design of wavelet filters $ \{\Hm_0,\Hm_1,\Gm_0,\Gm_1\}$ and 
% ii) design of downsampling function $\beta_{\Sc}$. 
% We propose a novel construction of two-channel wavelet 
% filterbanks for {\em any} undirected weighted graph which are 
% a) critically sampled and b) come with a perfect reconstruction 
% guarantee for all graph-signals.
%  We define a wavelet 
% filter to be an operator in $\RR^N$ domain which is localized 
% in both spatial and spectral domain of the graph.    
% Our contributions in this paper is to design a two-channel 
% wavelet filterbank on graphs which are a) critically sampled 
% b) can be applied to arbitrary undirected graph c) 
% with assured perfect reconstruction guarantee. 
For designing wavelet filters 
on graphs we exploit similar 
concepts of spectral decomposition 
as in \cite{Hammond'09}. Because of this, it is useful to define 
analysis wavelet filters $\Hm_0$ and $\Hm_1$ % In particular we choose 
in terms of spectral kernels $h_0(\lambda)$ and $h_1(\lambda)$ respectively.
%  for 
% . 
% and spectral kernels $g_0(\lambda)$ and $g_1(\lambda)$
% for synthesis wavelet filters $\Gm_0$ and $\Gm_1$ 
% we choose a low-pass/high-pass 
% spectral kernel pair $\{h_0(\lambda),h_1(\lambda)\}$ respectively. The low-pass kernel  $h_0(\lambda)$ has 
% non-zero d.c. response
% $h_0(0) > 0$ and attenuated response for high $\lambda$-values, whereas high-pass kernel $h_1(\lambda)$ has   
% a non-zero response $h_0(\lambda_N) > 0$ and attenuated response for low magnitude $\lambda$-values. 
Thus given the eigen-space decomposition of Laplacian matrix $\Lcb$ as in (\ref{eq:ESD_def}), the analysis filters can be represented as
\begin{equation}
 \begin{array}{l}
 \displaystyle {\Hm}_0 = h_0({\Lcb}) = \sum_{\lambda \in \sigma(G)} h_0(\lambda){\Pm}_{\lambda}   \\
 \displaystyle  {\Hm}_1 = h_1({\Lcb}) = \sum_{\lambda \in \sigma(G)} h_1(\lambda){\Pm}_{\lambda}
 \end{array}
\label{eq:spectral_tx}
\end{equation} 
\add[SN]{\protect Since the Laplacian matrix $\Lcb$ is real and symmetric, the filters designed in (\ref{eq:spectral_tx}) are also real and symmetric.} 
As described in Section~\ref{sec:downsampling}, 
% the downsampling in graphs can be explained with the help of 
% a downsampling function $\beta$ and corresponding downsampling 
% matrix $\Jm_{beta}$. From (\ref{eq:downsampling_spectral}), we observe
% that 
the spectral decomposition of the output of a $DU$ operation\remove[SN]{on graph}
with downsampling function $\beta$ yields a set of original signal 
coefficients  and a set of deformed signal coefficients which are generated 
by the projection of the original signal onto deformed eigenvectors. \add[SN]{In what follows,} 
% In the subsection below, 
we 
take the special case of 
% We first show that for an important 
% class of graphs 
% known as 
{\em bipartite graphs} for which \change[SN]{multiplication of 
downsampling matrix $\Jm_{\beta}$ with}{the $DU$ operation on} any eigenvector
produces an alias eigenvector at a mirror eigenvalue,
% (or eigenspace) of eigenvalue $\lambda$ 
% produces another eigenvector (or eigenspace) at a mirror eigenvalue
%  $(2-\lambda)$. 
% This 
% % leads to a 
% {\em spectrum folding} 
% phenomenon 
% around a central eigenvalue is 
% which 
a phenomenon, which is analogous to the ``aliasing'' effect observed in $DU$  
operations  in regular signal domain.
\add[SN]{ \protect This property of bipartite graphs, allows us to express the perfect reconstruction 
conditions for the two channel graph-filterbanks, as given in (\ref{eq:general_PR}), in simple terms.} 
Subsequently, we state necessary and sufficient 
conditions for a two-channel graph filter-bank, designed using spectral transforms, 
to provide aliasing-cancellation, perfect reconstruction 
and an orthogonal decomposition of any graph-signal and propose a solution similar to  
quadrature mirror 
filters (QMF) in regular signal domain which satisfies all of the above conditions. 
% We 
% abbreviated 
% as  design . 
For arbitrary graphs,
we formulate a
bipartite subgraph 
decomposition problem that
% The decomposition 
provides us with a disjoint collection of 
bipartite subgraphs whose union is $G$. 
A wavelet filterbank can be constructed 
on each  of these subgraphs
% is then used as a separate dimension 
% to filter and downsample. This 
leading to a 
multi-dimensional separable 
wavelet filterbank on any arbitrary graph. 
\add[SN]{Finally, we propose a multi-resolution implementation in which 
the proposed filterbanks 
can be recursively applied  to the downsampled output coefficients of each channel.}

\subsection{Downsampling in bipartite graphs}
\label{sec:fb_design_bpt}
A bipartite graph \change[SN]{$G = (\Sc_1,\Sc_2,\textit E)$}{$G = (L,H,\textit E)$} 
is a graph whose 
vertices can be divided into two disjoint 
sets \change[SN]{$\Sc_1$}{$L$} and \change[SN]{$\Sc_2$}{$H$}, such that 
every link connects a vertex 
in \change[SN]{$\Sc_1$}{$L$} to one in \change[SN]{$\Sc_2$}{$H$}. 
Bipartite graphs are 
also known as {\em two-colorable graphs}
since the 
vertices can be colored 
perfectly into two colors so that 
no two connected vertices 
are of the same color. 
% Bipartite 
% graphs have been found useful for in a large number of applications 
% such as for modelling matching problems and 
% decoding codewords in coding theory. 
Examples of bipartite graphs include 
tree graphs, cycle graphs and planar graphs with even degrees. 
\remove[SN]{In fact all 
regular 
signals in $N$-D space can be interpreted 
as situated on a regular 
lattice graph connected with $2N$ 
neighbors (left-right, up-down etc.) and hence are bipartite. By virtue of this property many characteristics 
of regular domain signals get easily translated to bipartite graphs. }
% 
% 
% This natural 
% partitioning of vertex set into two sets 
% intuitively 
% provides some evidence of favorable 
% properties of bipartite graphs 
% for  
% two-channel filterbanks design. Moreover, 
% the fact that all regularly sampled signals 
% can be interpreted as graph-signals  
% on bipartite regular lattice graphs
% lends more credence to our choice. 
In our analysis, we use the normalized form of the 
Laplacian matrix $\Lcb = \Dm^{-1/2}\Lm \Dm^{-1/2}$ 
for the bipartite graph \change[SN]{to avoid the non-uniformity in the 
connectivity of the nodes. Note that}{, which} in the case of regular graphs 
has the same 
set of eigenvectors as $\Lm$. \add[SN]{The normalization reweighs the edges of graph $G$ 
so that the degree of each node is equal to $1$.}
To understand the spectral interpretation of 
$DU$ operations in bipartite graphs, the
% \subsubsection{Downsampling in Bipartite Graphs}
following 
properties of bipartite graphs are useful:
%  in 
% our analysis:

\begin{lemma}[{\cite[Lemma 1.8]{fChung'97}}]
The following statements are equivalent for any graph $G$:
\begin{enumerate}
 \item $G$ is bipartite with bipartitions $H$ and $L$.
\item The spectrum of $\Lcb(G)$ is symmetric about  $1$ 
and the minimum and maximum eigenvalues of $\Lcb(G)$ are 
$0$ and $2$ respectively.
\item If $\uv = \begin{bmatrix}
            \uv_1^{T} & \uv_2^{T}
           \end{bmatrix}^T$ \change[SN]{be}{is} an eigenvector of $\Lcb$ with 
eigenvalue $\lambda$ with 
$\uv_1$ indexed on $H$ and $\uv_2$ indexed on $L$ (or vice-versa)
then the deformed eigenvector $\hat \uv = \begin{bmatrix}
            \uv_1^{T} & -\uv_2^{T}
           \end{bmatrix}^T$ is also an eigenvector of $\Lcb$ with eigenvalue $2
- \lambda$.
\end{enumerate}
\label{lem:lem1}
\end{lemma}
Based on these properties of bipartite graphs 
we state our key result below:
% For a bipartite graph, $G = (\Sc_1,\Sc_2,\textit E)$ a downsampling function can be defined 
% such that
% $\beta_n = 1$ if $n \in \Sc_1$ and 
% $\beta_n = -1$ if $n \in \Sc_2$. This leads to following result:
\begin{proposition}
Given a bipartite graph $G = (L,H,\textit E)$ with Laplacian matrix $\Lcb$, if we choose downsampling 
function $\beta$ as $\beta_{H}$ or $\beta_{L}$ as defined in (\ref{eq:beta_gen}), and
if $\Pm_{\lambda}$
is the eigen-space corresponding to the eigenvalue $\lambda_{s}$ 
then 
\begin{equation}
\Jm_{\beta}\Pm_{\lambda} = \Pm_{2 - \lambda}\Jm_{\beta}.
 \label{eq:spectral_folding}
\end{equation}
Alternatively, 
if $\uv_{\lambda}$
is an eigen-vector of $\Lcb$ 
with eigenvalue $\lambda$ then $\Jm_{\beta}\uv_{\lambda}$
is also an eigen-vector of $\Lcb$ with eigen-value $2 -\lambda$. 
%  (or  $ since 
% $\Jm_{\beta}^2 = \Id$).
% is also an eigen-vector of $\Lcb$ with eigen-value $2 -\lambda_{s} = \lambda_{N-s}$.
 \label{prop:prop1}
\end{proposition}
\begin{IEEEproof}
% \begin{itemize}
% \item[]Note that since  $\Jm_{\beta}= diag\{\pm 1\}$~~$\rightarrow$ ~~ $\Jm_{\beta}^2 = \Id$ (identity matrix), 
% the claim in Proposition~\ref{prop:prop1} can also be written as
% to $\Jm_{\beta}\Pm_{\lambda} = \Pm_{2 - \lambda}\Jm_{\beta}$.
% \item[] 
% Note that since $\Jm_{\beta}^2 = \Id$, the result can also be written as $\Jm_{\beta}\Pm_{\lambda}\Jm_{\beta} = \Pm_{2 - \lambda}$.
Let $\lambda$ be an eigenvalue of $G$ with multiplicity $k$. 
This implies that there exists an orthogonal set of $k$ eigenvectors 
$\{\uv_i\}_{\lambda_i = \lambda}$ of Laplacian matrix $\Lcb$ with eigenvalue $\lambda$. 
The projection matrix $\Pm_{\lambda}$ corresponding to  
$\lambda$ is given by $\Pm_{\lambda} = \sum_{\lambda_i = \lambda}\uv_i.\uv_i^t$. 
Note that in case of $k > 1$, the eigenspace $\Pm_{\lambda}$ is still unique 
whereas the eigenvectors $\{\uv_i\}_{\lambda_i = \lambda}$ are only 
unique up to a unitary transformation.
% \item[] 
If 
the downsampling function $\beta$ is chosen as $\beta_H$ or $\beta_L$, 
then the deformed eigenvector $\hat \uv$ in Lemma \ref{lem:lem1} is equal to $\Jm_{\beta} \uv$, which 
is an eigen-vector of $\Lcb$ with eigen-value $2 -\lambda$. 
It can also be seen that if eigenvectors $\{\uv_i\}_{\lambda_i = \lambda}$ are orthogonal to each other then 
so are the deformed set of eigenvectors $\{\Jm_{\beta}\uv_i\}_{\lambda_i = \lambda}$ and form basis of eigenspace $\Pm_{2 - \lambda}$.
% It can be seen that if $\uv_i = \begin{bmatrix}
%             \uv_{i,1}^{t} & \uv_{i,2}^{t}
%            \end{bmatrix}^T$ is an eigenvector of $\Lcb$ with 
% eigenvalue $\lambda$ with 
% $\uv_{i,1}$ indexed on $\Sc_1$ and $\uv_{i,2}$ indexed on $\Sc_2$
% then $\Jm_{\beta}\uv_i = \begin{bmatrix}
%             \uv_{i,1}^{t} & -\uv_{i,2}^{t}
%            \end{bmatrix}^T$ is also an eigenvector of $\Lcb$ with eigenvalue $2 - \lambda$ from Theorem ~\ref{thm:thm1} \\
% \item[] 
% \item[] 
Therefore, $\Lcb \Jm_{\beta}\Pm_{\lambda}\Jm_{\beta} = \sum_{\lambda_i = \lambda}\Lcb .\Jm_{\beta}\uv_i.(\Jm_{\beta}\uv_i)^t = \sum_{\lambda_i = \lambda}(2-\lambda).\Jm_{\beta}\uv_i.(\Jm_{\beta}\uv_i)^t = (2-\lambda)\Pm_{2 - \lambda}$, therefore
$\Jm_{\beta}\Pm_{\lambda}\Jm_{\beta} = \Pm_{2 - \lambda}$ which implies that  $\Jm_{\beta}\Pm_{\lambda} = \Pm_{2 - \lambda}\Jm_{\beta}$.
% \end{itemize}
% Since  $\Jm_{\beta}= diag\{\pm 1\}$~~$\rightarrow$ ~~ $\Jm_{\beta}^2 = \Id$ (identity matrix). Thus it will suffice to prove 
% that $\Jm_{\beta}\Pm_{\lambda} = \Pm_{2 - \lambda}\Jm_{\beta}$. \\
% $\Rightarrow$~~ 
% 
%  and the result follows from Theorem~\ref{thm:thm1}.
\end{IEEEproof}
\change[SN]{The phenomenon is termed  as}{We term this phenomenon, }{\em spectrum folding} in bipartite graphs, as the deformed eigenvector (or eigenspace) 
for any $\lambda \in \sigma(G)$ appears as another eigenvector (or eigenspace) at a mirror eigenvalue around $\lambda = 1$. 
To understand it, let $\fv$ be an $N$-D graph-signal on bipartite graph $G = (L,H,\textit E)$ with eigenspace decomposition
\begin{equation}
\fv = \sum_{\lambda \in \sigma(G)}\Pm_{\lambda}\fv = \sum_{\lambda \in \sigma(G)}\fv^{\lambda},
 \label{eq:eigenspace_decomp}
\end{equation}
where $\fv^{\lambda} = \Pm_{\lambda}\fv$ is the projection of $\fv$ onto the eigenspace $V_{\lambda}$ and let the 
output signal after $DU$ operation with downsampling function $\beta_{L}$ (or $\beta_{H}$) be $\fv_{du}$. Then the 
$V_{\lambda}$ eigenspace projection of the output signal is given as:
\begin{equation}
 \fv_{du}^{\lambda} = \Pm_{\lambda}\fv_{du} = \frac{1}{2}\Pm_{\lambda}\fv + \Pm_{\lambda}\Jm_{\beta_L}\fv, 
\end{equation}
which using (\ref{eq:spectral_folding}), can be written as:
\begin{eqnarray}
 \fv_{du}^{\lambda} & = & \frac{1}{2}\Pm_{\lambda}\fv + \Jm_{\beta_L}\Pm_{2- \lambda}\fv \nonumber \\
& = & \frac{1}{2}(\fv^{\lambda} + \Jm_{\beta_L}\fv^{2 - \lambda}). 
\label{eq:aliasing_def}
\end{eqnarray}
In (\ref{eq:aliasing_def}), the distortion term $\Jm_{\beta_L}\fv^{2 - \lambda}$, which arises due to the 
downsampling of $\fv^{\lambda}$ has the same coefficients as 
that of $\fv^{2 - \lambda}$ (except for different signs). 
Further, the eigenspace decomposition 
of the output signal can be written as:
\begin{equation}
 \fv_{du} = \frac{1}{2}\sum_{\lambda \in \sigma(G)} (\fv^{\lambda} + \Jm_{\beta}\fv^{2 - \lambda}) = \frac{1}{2} (\fv + \fv^{alias})
\end{equation}
% Thus the output $\xv_{du}$ can be written as 
% \begin{equation}
%  \xv_{du}= \frac{1}{2}\Pm_{\lambda}(\Id+\Jm_{\beta})\xv = \frac{1}{2}(\Pm_{\lambda}+ \Jm_{\beta}\Pm_{2-\lambda})\xv = \frac{1}{2}(\xv^{\lambda} + \Jm_{\beta}\xv^{2 - \lambda})
% \end{equation}
{\em In other words, the output signal is the average of the original signal and a shifted and aliased version of the original signal, and 
\add[SN]{hence the term spectral folding.} In the next Section, we utilize this property to design perfect reconstruction filterbanks for bipartite graphs.} 
\subsection{Two-Channel Filterbank Conditions for Bipartite Graphs}
% In this section we use the result of Proposition~\ref{prop:prop1} to design QMF filterbanks on bipartite graphs. 
Referring again to Figure 
\ref{fig:graph_filterbank}, for bipartite graph \change[SN]{$G = (\Sc_1,\Sc_2,\textit E)$}{$G = (L,H,\textit E)$}, let 
\change[SN]{$\beta = \beta_{\Sc_1}$}{$\beta_{H}  = \beta$} 
be the downsampling function for $\Hm_1$ filter channel and 
\change[SN]{$\beta_{\Sc_2}= -\beta$}{$\beta_{L} = -\beta$} be the downsampling function for $\Hm_0$ channel. 
Thus the nodes in \change[SN]{$\Sc_1$}{$H$} 
only retain the output of \change[SN]{transform $\Hm_0$}{highpass channel} and nodes in \change[SN]{$\Sc_2$}{$L$} 
retain the output of \change[SN]{transform $\Hm_1$}{the lowpass channel}. \remove[SN]{only and the overall output is critically sampled.}
% The reconstructed signal $\hat \fv$ can be written in terms of original signal as
% \begin{eqnarray}
%  \displaystyle \hat \fv &=& \frac{1}{2}\left(\Gm_0(\Id-\Jm_{\beta})\Hm_0 +\Gm_1(\Id + \Jm_{\beta})\Hm_1\right)\xv \nonumber \\ 
%   \displaystyle &=& \frac{1}{2}(\Gm_0\Hm_0+ \Gm_1\Hm_1)\xv + \frac{1}{2}(\Gm_1\Jm_{\beta}\Hm_1 - \Gm_0\Jm_{\beta}\Hm_0  )\xv  \nonumber \\ 
% \displaystyle &=&\frac{1}{2}\Tm_{eq}\xv + \frac{1}{2}\Tm_{alias}\xv
% \label{eq:equiv_tx}
% \end{eqnarray}
% where $\Tm_{eq}$ and $\Tm_{alias}$ are defined in (\ref{eq:general_PR}). 
% where the 
% For anti-aliasing and for perfect reconstruction 
In our proposed design, we also choose the synthesis filters $\Gm_0$ and $\Gm_1$ to be spectral filters with kernels $g_0(\lambda)$ and $g_1(\lambda)$ respectively
\footnote[3]{In general, synthesis filters do not have to be based on the spectral design. 
A case is presented in our previous work~\cite{ICIP'10} with linear kernel spectral analysis filters and non-spectral synthesis filters. }.
Then,
by using~(\ref{eq:eigenspace_prop1}) and~(\ref{eq:spectral_tx}) the perfect reconstruction conditions in~(\ref{eq:general_PR}) can be rewritten as:
\begin{eqnarray}
\displaystyle \Tm_{eq} & = & \Gm_0\Hm_0+ \Gm_1\Hm_1 \nonumber \\
&=& \sum_{\lambda \in \sigma(G)}\left(g_0(\lambda)h_0(\lambda) + g_1(\lambda)h_1(\lambda)\right)\Pm_{\lambda} \nonumber \\
\displaystyle \Tm_{alias} &=& \Gm_1\Jm_{\beta}\Hm_1- \Gm_0\Jm_{\beta}\Hm_0  \nonumber \\
&=& \sum_{\lambda, \gamma \in \sigma(G)}\left(g_1(\lambda)h_1(\gamma) - g_0(\lambda)h_0(\gamma)\right)\Pm_{\lambda}\Jm_{\beta}\Pm_{\gamma}.
\label{eq:T_eq}
\end{eqnarray}
% with the result of Proposition~\ref{prop:prop1} we obtain:
% \begin{eqnarray}
% \label{eq:T_alias}
% \end{eqnarray}
\subsubsection{Aliasing cancellation}
 Using~(\ref{eq:eigenspace_prop1}) and the spectral folding property of bipartite graphs in (\ref{eq:spectral_folding}), $\Tm_{alias}\fv$ 
can be written as:
\begin{eqnarray}
% \begin{array}{l}
%  \displaystyle \Tm_{eq} = \Gm_0^t\Hm_0+ \Gm_1^t\Hm_1 = \sum_{\lambda \in \sigma(G)}\left(g_0(\lambda)h_0(\lambda) + g_1(\lambda)h_1(\lambda)\right)\Pm_{\lambda} \\ \nonumber
\displaystyle \Tm_{alias}\fv &=& \sum_{\lambda \in \sigma(G)}\left(g_1(\lambda)h_1(2 - \lambda) - g_0(\lambda)h_0(2 - \lambda)\right)\Jm_{\beta}\Pm_{2 -\lambda}\fv \nonumber \\
\displaystyle  &=& \sum_{\lambda \in \sigma(G)}\left(g_1(\lambda)h_1(2 - \lambda) - g_0(\lambda)h_0(2 - \lambda)\right)\Pm_{\lambda}\Jm_{\beta}\fv^{2 -\lambda}
% \sum_{\lambda}\sum_{\gamma}\left(g_1(\lambda)h_1(\gamma) - g_0(\lambda)h_0(\gamma)\right)\Pm_{\lambda}\Pm_{2 - \gamma}\Jm_{\beta} \nonumber \\
% \displaystyle 
% \end{array}
\label{eq:T_alias}
\end{eqnarray}
Since, $\Jm_{\beta}\fv^{2-\lambda}$ is the aliasing term corresponding to $\fv^{\lambda}$,
$\Tm_{alias}\fv$ is the aliasing part of the reconstructed signal, and an
{\em alias-free reconstruction using spectral filters
is possible if and only if} for all $\lambda$ in $\sigma(G)$, 
% 
% on input-signal $\fv$ is equivalent to the filtering operation of a graph-filter with spectral kernel 
% 
%  
% operates on the aliasing term with spectral kernel $g_1(\lambda)h_1(2-\lambda) - g_0(\lambda)h_0(2-\lambda)$. 
% therefore 
% {\em aliasing cancellation}  is
\begin{equation}
g_0(\lambda)h_0(2-\lambda) - g_1(\lambda)h_1(2-\lambda) = 0.
 \label{eq:aliasing_cancel}
\end{equation}
\subsubsection{Perfect reconstruction}
Perfect reconstruction means that the reconstructed signal $\hat \fv$ is the same as (or possibly a scaled version of) 
the input signal $\fv$.
 $\Tm_{eq} + \Tm_{alias} = \Id$. Therefore assuming the filterbanks cancel aliasing, {\em the perfect reconstruction 
can be obtained if and only if} $\Tm_{eq} = c^2\Id$ for some scalar constant $c$.
% are 
% 
% By combining~(\ref{eq:T_alias}) and our result $\Pm_{\lambda}\Jm_{\beta} = \Jm_{\beta}\Pm_{2 - \lambda}$ we obtain:
% \begin{equation}
% \Tm_a = \frac{1}{2}\sum_{\lambda \in \sigma(G)}\underbrace{\left [(g_0(\lambda)h_0(\lambda) + g_1(\lambda)h_1(\lambda))\Id + (h_1(\lambda)g_1(2-\lambda) - h_0(\lambda)g_0(2-\lambda))\Jm_{\beta}\right]}_{\Dm_{\lambda}}\Pm_{\lambda} 
%  \label{eq:T_eq1}
% \end{equation}
% 
% For perfect reconstruction $\Dm_{\lambda} = c^2\Id$ which implies that the term corresponding to $\Jm_{\beta}$ should be zero and the term corresponding to $\Id$ 
% should be constant for all graph-frequencies $\lambda$.
% % Similarly the non-aliasing part of transform is $\Tm_{eq}$ and for perfect reconstruction $\Tm_{eq}$ should have a constant 
% % response at all eigenspaces $\Pm_{\lambda}$. 
Thus, a necessary and sufficient condition for {\em perfect reconstruction}, using spectral filters, in bipartite graphs filterbanks is 
that for all $\lambda$ in $\sigma(G)$,
\begin{eqnarray}
g_0(\lambda)h_0(\lambda) + g_1(\lambda)h_1(\lambda) = c^2, \nonumber \\
g_0(\lambda)h_0(2-\lambda) - g_1(\lambda)h_1(2-\lambda) = 0.
 \label{eq:perfect_reconstruct}
\end{eqnarray}
\subsubsection{Orthogonality}
The equivalent analysis filter $\Tm_{a}$ in the filterbank 
of Figure~\ref{fig:graph_filterbank} is given as 
\begin{eqnarray}
\Tm_{a} &=& \frac{1}{2}\left((\Id-\Jm_{\beta})\Hm_0 +(\Id + \Jm_{\beta})\Hm_1\right) \nonumber \\ 
& = & \frac{1}{2}(\Hm_0 + \Hm_1) + \frac{1}{2}\Jm_{\beta}(\Hm_1 - \Hm_0)
 \label{eq:analysis_tx}
\end{eqnarray}
The filterbank provides an {\em orthogonal decomposition} of the graph signal if $\Tm_{a}^{-1} = \Tm_{a}^t$, 
which implies  $\Tm_a\Tm_a^t = \Tm_a^t\Tm_a = \Id$. 
Since, the spectral filters as well as the downsampling matrix $\Jm_{\beta}$ are symmetric, 
$\Tm_a^t\Tm_a$ can be expanded as: 
% &=& \left(\frac{1}{2}(\Hm_0 + \Hm_1) + \frac{1}{2}(\Hm_1 - \Hm_0)\Jm_{\beta}\right)\left(\frac{1}{2}(\Hm_0 + \Hm_1) + \frac{1}{2}\Jm_{\beta}(\Hm_1 - \Hm_0)\right) \\ \nonumber
\begin{equation}
\Tm_{a}^t\Tm_a =  \frac{1}{2}\left(\Hm_0^2+\Hm_1^2 + \Hm_1\Jm_{\beta}\Hm_1 -\Hm_0\Jm_{\beta}\Hm_0 \right) 
\label{eq:tx_orthogonal1}
\end{equation}
Combining (\ref{eq:spectral_tx}) and (\ref{eq:tx_orthogonal1}) we obtain:
\begin{eqnarray}
\Tm_{a}^t\Tm_a & = & 1/2\sum_{\lambda \in \sigma(G)}\underbrace{(h_0^2(\lambda)+h_1^2(\lambda))}_{C_{\lambda}}\Pm_{\lambda} \nonumber \\
& + & 1/2 \sum_{\lambda \in \sigma(G)}\underbrace{(h_1(\lambda)h_1(2 - \lambda)-h_0(\lambda)h_0(2 - \lambda))}_{D_{\lambda}}\Jm_{\beta}\Pm_{\lambda}
\label{eq:tx_orthogonal2}
\end{eqnarray}
% & = & \frac{1}{2}\sum_{\lambda \in \sigma(G)}\underbrace{\left [(h_0^2(\lambda)+h_1^2(\lambda))\Id + (h_1(\lambda)h_1(2 - \lambda)-h_0(\lambda)h_0(2 - \lambda))\Jm_{\beta}\right]}_{\Dm_{\lambda}}\Pm_{\lambda} 
% 
% 
% 
% 
% 
Thus, orthogonality can be obtained if and only if $C_{\lambda}\Id + D_{\lambda}\Jm_{\beta}  = c^2\Id$ for some constant $c$
and for all $\lambda \in \sigma(G)$, which
is possible if and only 
if $D_{\lambda} = 0$ and $C_{\lambda} = c^2$ for 
all $\lambda$. Thus, a necessary and sufficient condition 
for orthogonality in bipartite graph filterbanks using spectral filters is :
\begin{eqnarray}
h_0(\lambda)h_0(2 - \lambda)-h_1(\lambda)h_1(2 - \lambda) &= & 0  \nonumber  \\
 h_0^2(\lambda)+h_1^2(\lambda) & = & c^2.
\label{eq:orthogonality} 
\end{eqnarray}
\add[SN]{\protect Note that, comparing (\ref{eq:perfect_reconstruct}) and~(\ref{eq:orthogonality}), 
the orthogonality conditions can be obtained from the perfect reconstruction conditions
by selecting
$g_0(\lambda) = h_0(\lambda)$ and $g_1(\lambda) = h_1(\lambda)$. This is analogous to the case of 
standard filterbanks and 
leads to our proposed graph-QMF design as explained in the next Section.}
% Similarly 
% \begin{eqnarray}
% \Tm_{a}\Tm_a^t &=& \left(\frac{1}{2}(\Hm_0 + \Hm_1) + \frac{1}{2}\Jm_{\beta}(\Hm_0 - \Hm_1)\right)\left(\frac{1}{2}(\Hm_0 + \Hm_1) + \frac{1}{2}(\Hm_0 - \Hm_1)\Jm_{\beta}\right) \\ \nonumber
% & = & \frac{1}{4}\sum_{\lambda \in \sigma(G)}((h_0(\lambda)+h_1(\lambda))^2 + (h_0(2 - \lambda)-h_1(2 -\lambda))^2)\Pm_{\lambda} \\ \nonumber
% &+& \frac{1}{4}\sum_{\lambda \in \sigma(G)}((h_0^2(\lambda)+h_0^2(2 -\lambda)) - (h_1^2(\lambda)+h_1^2(2 -\lambda)) )\Jm_{\beta}\Pm_{\lambda}
% \label{eq:tx_orthogonal2}
% \end{eqnarray}
\subsection{Proposed Solution: Graph-QMF Design}
\label{sec:graph-QMF}
\begin{figure*}[htb]
\centering
\includegraphics[width=6in]{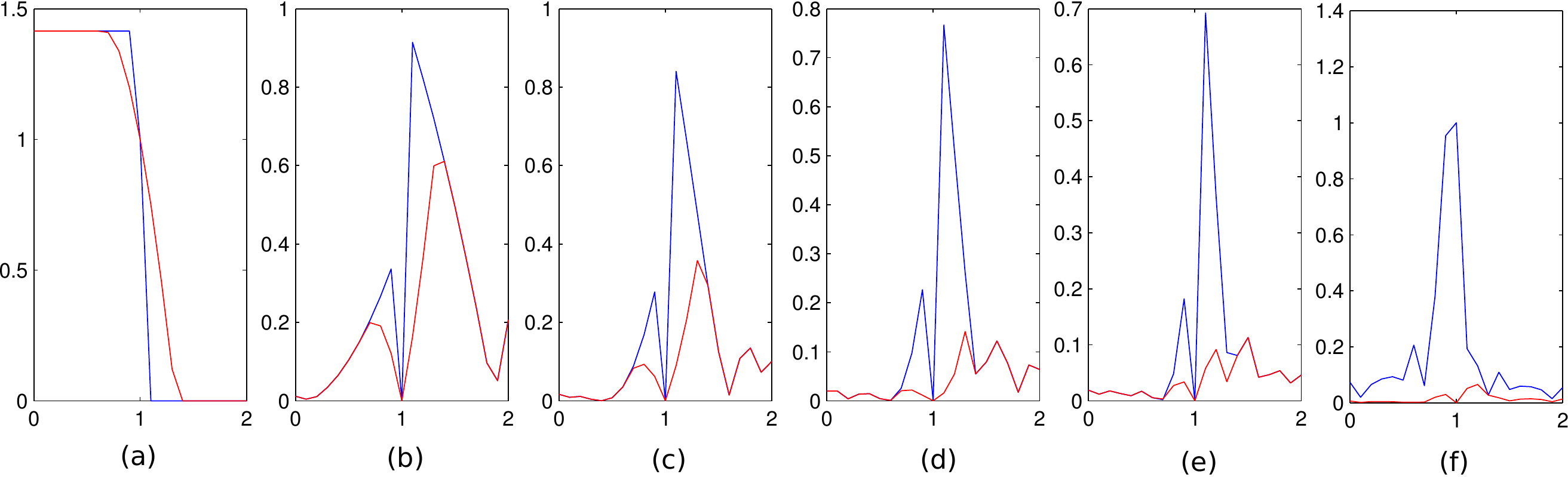}
\caption[Polynomial Approximations of desired ideal filters]{\footnotesize
(a) Ideal kernel (blue) vs. Meyer's wavelet kernel (red). It can be seen that Meyer's wavelet 
has smoother 
transition 
% in response 
at $\lambda =1$ than the ideal kernel, (b)-(f) the reconstruction error magnitudes 
between original kernels and their polynomial approximations of order $2,4,6,8$ and $10$ respectively:
ideal kernel (blue curves) and Meyers kernel(red curve).}
\label{fig:ideal_filt_approx}%
\end{figure*}

We extend the 
well-known quadrature mirror filter (QMF) solution to the case of bipartite graphs. 
% which 
% satisfies the aliasing cancellation condition given in (\ref{eq:aliasing_cancel}). 
Our proposed solution, termed as graph-QMF, leads to the design of a single spectral kernel $h_0(\lambda)$ 
by selecting the other spectral kernels as:
\begin{equation}
 \begin{array}{l}
h_1(\lambda) = h_0(2-\lambda) \\ 
g_0(\lambda) = h_0(\lambda) \\ 
g_1(\lambda) = h_1(\lambda) = h_0(2-\lambda)    
\end{array}
\label{eq:graph-QMF}
\end{equation}
% proposed 
\begin{proposition}[QMF Filters on Graph]
For a bipartite graph \change[SN]{$G= (\Sc_1,\Sc_2,\textit E)$}{$G= (L,H,\textit E)$}, 
let a two-channel filterbank be as shown in Figure~\ref{fig:graph_filterbank} with 
the downsampling function \change[SN]{$\beta = \beta_{\Sc_1}$}{$\beta = \beta_H$} 
and with spectral filters
$\{\Hm_0,\Hm_1,\Gm_0,\Gm_1\}$ corresponding to spectral kernels $\{h_0(\lambda),h_1(\lambda),g_0(\lambda),g_1(\lambda)\}$ 
respectively. Then for 
any arbitrary choice of kernel $h_0(\lambda)$, the proposed graph-QMF solution     
cancels aliasing in the filterbank. In addition for $h_0(\lambda)^2 + h_0(2 - \lambda)^2 = c^2$ for all $\lambda \in \sigma(G)$ and $c \neq 0$ the 
filterbank provides perfect reconstruction and an orthogonal decomposition of graph-signals.
% \begin{equation}
%  \begin{array}{l}
%  \\ \nonumber
%    \\ \nonumber
%   g_0(\lambda) = h(\lambda) \\ \nonumber
%   g_1(\lambda) = h(2-\lambda) 
%  \end{array}
% \label{eq:QMF_filters}
% \end{equation}
 \label{prop:prop2}
\end{proposition}
\begin{IEEEproof}
% the overall filterbank response in the $\lambda$-eigenspace is given as $h^2(\lambda) + h^2(2-\lambda)$ which should be bounded away 
% from zero for a loss-less recovery of the original signal. 
% \begin{itemize}
Substituting (\ref{eq:graph-QMF}) into~(\ref{eq:aliasing_cancel}) leads to $g_0(\lambda)h_0(2-\lambda) - g_1(\lambda)h_1(2-\lambda) = 0$ and aliasing is indeed 
canceled. The reconstructed signal $\hat \xv$ in this case is  simply equal to $(1/2)\Tm_{eq}\xv$ and can be written as:
\begin{equation}
 \hat \xv = \frac{1}{2}\sum_{\lambda \in \sigma(G)}(h^2(\lambda) + h^2(2-\lambda))\xv^{\lambda}
\end{equation}
Thus for $(h^2(\lambda) + h^2(2-\lambda)) = c^2$ and $c \neq 0$, the reconstructed signal $\hat \xv = \frac{c^2}{2} \xv$ is a scaled version of original signal.
Similarly applying the mirror design $h_1(\lambda) = h_0(2 - \lambda)$ in the conditions (\ref{eq:orthogonality}) 
we get   $h_0(\lambda)h_0(2 - \lambda)-h_1(\lambda)h_1(2 - \lambda) = 0$ and $h_0^2(\lambda) + h_1^2(\lambda) = c^2$ 
and hence 
% , 
% we get 
% \begin{equation}
% \Tm_a\Tm_a^t = \Tm_a^t\Tm_a = \frac{1}{2}\sum_{\lambda \in \sigma(G)}(h_0^2(\lambda)+h_0^2(2-\lambda))\Pm_{\lambda}
% \label{eq:tx_orthogonal3}
% \end{equation}
% \item[] Thus by applying the second condition $h_0^2(\lambda) + h_0^2(2 - \lambda) = c^2$ we get $\Tm_{a}\Tm_a^t = (c^2/2)\Id$ and 
corresponding 
analysis side transform $\Tm_{a}$ is orthogonal. 
%  
% and $h_0^2(\lambda) + h_0^2(2 - \lambda) = c^2$ we get $\Tm_{a}^t\Tm_a = (c^2/2)\Id$ 
\end{IEEEproof}
% \subsubsection{Spectral kernel design}
\change[SN]{\protect According to Proposition~\ref{prop:prop2}, for designing a 
two-channel orthogonal filterbank on bipartite graphs, using spectral filters, 
% with all above mentioned 
% properties,
% at minimum 
we require
a kernel $h_0(\lambda)$ which 
% also needs 
% to 
satisfies the 
constraints $h_0^2(\lambda) + h_0^2(2-\lambda) = c^2 $ 
for all $\lambda$ in the spectrum of the graph $G$.}{ \protect We now consider the design 
of kernels $h_0(\lambda)$ satisfying the design constraint of Proposition \ref{prop:prop2}, i.e., 
for which $h_0^2(\lambda) + h_0^2(2-\lambda) = c^2 $ for all $\lambda \in \sigma(G)$.}
% The remaining kernels
% can be computed from this kernel according to (\ref{eq:graph-QMF}).
% Additionally, 
% we need the kernel $h_0(\lambda)$ to be localized over graph-frequencies 
% $\lambda \leq 1$.
For maximum spectrum splitting in the 
two channels of the filterbank,
the ideal choice of kernel $h_0(\lambda)$ 
% in spectral domain 
would be a lowpass rectangular function on $\lambda$ 
given as:
\begin{equation}
\displaystyle h_0^{ideal}(\lambda) =
\left\{
 \begin{array}{ll}
c & \mbox{if } \lambda < 1 \\
c/\sqrt(2) & \mbox{if } \lambda = 1 \\
0 & \mbox{if } \lambda > 1
\end{array}
\right.
\label{eq:h_ideal}
\end{equation}
The
corresponding ideal filter is given by 
\begin{equation}
\Hm_{0}^{ideal} = \sum_{\lambda < 1} c\Pm_{\lambda} + \frac{c}{\sqrt{2}}\Pm_{\lambda = 1}
\label{eq:ideal_filt}
\end{equation}
Note that the ideal transform has a non-analytic spectral kernel response with sharp peaks and is therefore 
a 
global transform (i.e., the filter operations are not localized). % well localized in space. 
Even analytic 
solutions of the constraint equation  
$h_0^2(\lambda) + h_0^2(2-\lambda) = c^2$,
such as 
$h_0(\lambda) = c\sqrt{1-\lambda/2}$ or $h_0(\lambda) = c~cos(\pi\lambda/4)$, 
are not very well localized in the spatial domain.
% On the other hand 
By relaxing the constraints 
% $h_0^2(\lambda) + h_0^2(2-\lambda) = c^2$, 
one can obtain spatially localized solutions 
at the cost 
of some 
small reconstruction error and near-perfect 
orthogonality.
One such solution is the approximation of the 
desired kernel with a polynomial kernel.
% \footnote[4]{Another approach is to use the QMF filter design method proposed by Johnston 
% \cite{Johnston} which defines an objective function based on desired stop-band attenuation 
% error of $h(\lambda)$ and reconstruction error. }  
% It has been shown that 
% However the resulting ideal filters would be 
% For localization in space, 
% one for which the resulting 
% filterbank is well localized 
% in space and concentrated over low-magnitude eigenvalues (i.e. $\lambda \leq 1$). In particular 
We 
% specifically 
choose polynomial approximations of the desired 
kernel due to the following localization property for corresponding transforms: 
% polynomials of the Laplacian matrix:
\begin{lemma}[\cite{Hammond'09}]
 Let $h_0(\lambda)$ be a polynomial of degree $k$ and let $\Lcb$ be the normalized 
Laplacian matrix for any weighted graph $G$, then the matrix polynomial $\Hm_0 = h_0(\Lcb)$ is exactly 
$k$-hop localized at each node of $G$. In other words for any two nodes $n$ and $m$ if $m \notin \Nc_k(n)$ then $\Hm_0(n,m) = 0$.  
\label{lem:polynomial_kernel}
\end{lemma}
Further, we choose a minimax polynomial 
approximation which 
minimizes 
the Chebychev norm (worst-case norm) of the 
reconstruction error since it has been shown 
in~\cite{Hammond'09} that it  
% A common approach 
% The polynomial 
% approximation of 
% the desired 
% kernel  and lead to 
also minimizes the upper-bound 
on the error $||H^{ideal}- H^{poly}||$ between ideal  
and approximated filters.  
Thus, in order to localize \change[SN]{it}{the filters} on \add[SN]{the} graph, we 
% approximate 
% 
%  but is not localized in space. 
% 
approximate $h_0^{ideal}$ with \add[SN]{the} 
truncated Chebychev polynomials \add[SN]{(which are a good approximation of minimax polynomials)} of 
different orders. 
However 
since $h_0^{ideal}$ is a rectangular function it projects a lot of its 
energy in 
the truncated part of the polynomial expansions and as a result the 
polynomial approximation errors for $h_0^{ideal}$ are high. 
A possible solution of this problem 
is to soften the ideal case, by finding a smooth function that is 
low-pass and 
satisfies the constraint. An analogous construction in regular signal processing 
is  {\em Meyer's wavelet} design which replaces the brick-wall type ideal 
frequency-response with a smooth scaling function that satisfies the orthogonality 
and scaling requirements. By a change in variable from $\omega \in [-1, 1]$ to 
$\lambda \in  [0 ,2]$ we can extend Meyer's wavelet construction in the case of 
bipartite graph. The construction involves choosing a function $\nu(x)$ such that 
$\nu(\lambda) = 0 $ for $\lambda \leq 0$ , $\nu(\lambda) = 1 $ for $\lambda \geq 1$ and $\nu(\lambda) + \nu(1-\lambda) = 1$
everywhere. One such function is given as: 
% given as 
%  \item[a.] Choose a function $\nu(x)$ as
\begin{equation}
\displaystyle \nu(\lambda) =
\left\{
 \begin{array}{ll}
1 & \mbox{if } \lambda \leq 0\\
3\lambda^2 - 2\lambda^3 & \mbox{if } 0 \leq \lambda \leq 1 \\
0 & \mbox{if } 1 \geq \lambda
\end{array}
\right.
\label{eq:nu_x}
\end{equation} 
The smooth kernel is then given as:
 \begin{equation}
\displaystyle h_0^{Meyer}(\lambda) = \sqrt{\nu(2 - \frac{3}{2}\lambda)} 
\label{eq:meyer_x}
\end{equation} 
In Figure~\ref{fig:ideal_filt_approx}(a), we 
plot the ideal and Meyer wavelet 
kernels and in Figures~\ref{fig:ideal_filt_approx}(b)-(f) we plot  
the reconstruction errors between desired kernels and 
their polynomial 
approximations of different orders. 
It can be seen that Meyer's wavelet 
approximations yield small reconstruction 
errors as compared to 
% (red plots) 
% are smoother than 
ideal-filter 
approximations. Thus by choosing $h(\lambda)$ as the low-order polynomial 
approximations of smooth low-pass functions (such as Meyer's wavelets ), 
we obtain near perfect reconstruction QMF wavelet filters on any bipartite 
graph which are very well localized in spatial domain. 
  
% We provide the explicit construction of 
% these polynomial kernel wavelet-filters in Section~\ref{sec:experiments}. 
% In Figure \ref{fig:poly_approx} 
% we show some low order polynomial approximations of a desirable low-pass kernel. 

% \change[SN]{ \protect \subsection{Decomposition of a Graph Into Bipartite Subgraphs}}
% {\protect 
%  }
\subsection{Multi-dimensional separable wavelet filterbanks for arbitrary graphs}
\label{sec:bpt_decomp}
\begin{figure*}[htb]
\centering
\includegraphics[width=5in]{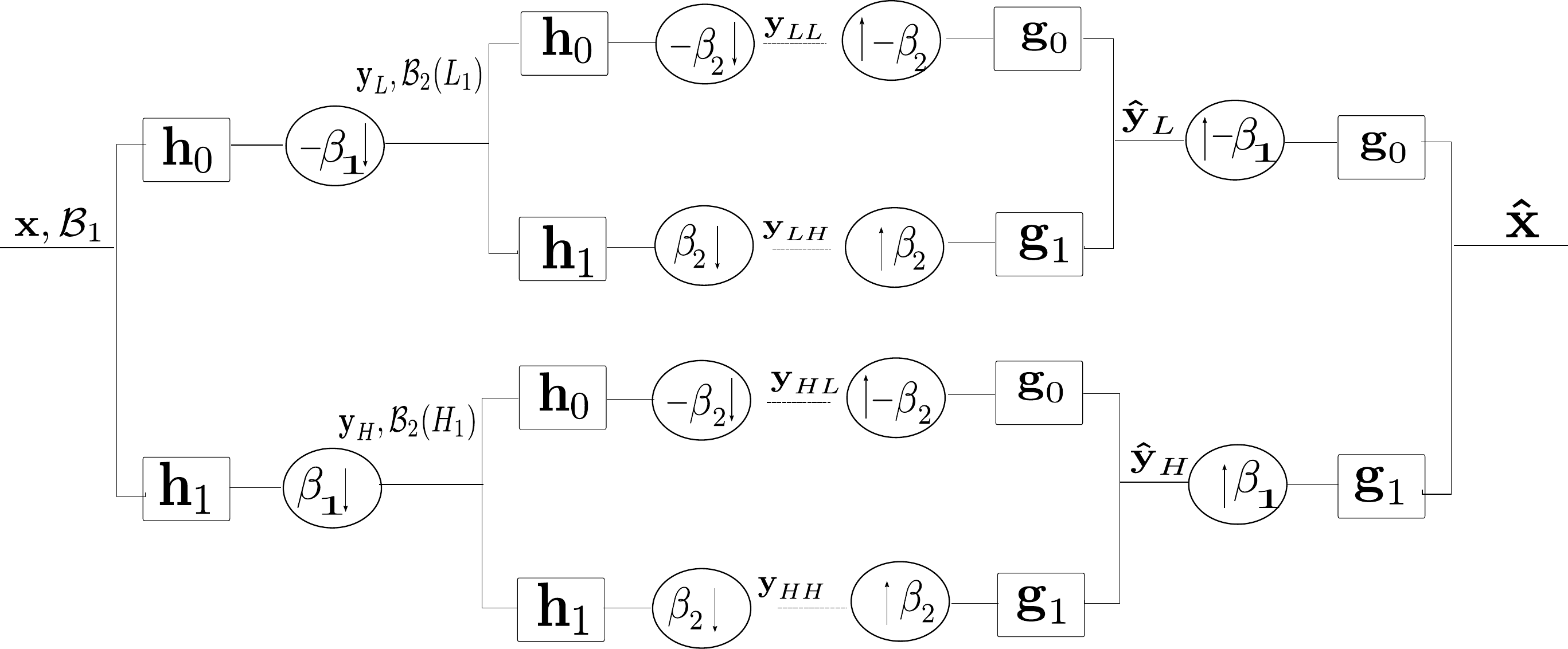}
\caption[A 2D Separable two-channel Filter Bank ]{\scriptsize Block diagram of a 2D Separable two-channel Filter Bank: 
the graph $G$ is first decomposed into two bipartite subgraphs $\Bc_1$ and $\Bc_2$, using the proposed decomposition scheme. 
The first two-channel filterbank is  designed on $\Bc_1$.
The filtering and downsampling on $\Bc_1$
creates output coefficients $\yv_H$ and $\yv_L$, stored on the sets $H_1$ and $L_1$, respectively. 
The second filterbank is designed on $\Bc_2$, which operates separately on signals $\yv_H$ and $\yv_L$ using the links 
of bipartite subgraphs $\Bc_{2}(H_1)$ and 
$\Bc_{2}(L_1)$ respectively. This creates $4$ sets of output transform coefficients, denoted as 
$\yv_{HH}, \yv_{HL}, \yv_{LH}$ and,$\yv_{LL}$, which are stored at disjoint sets of nodes, 
given as $H_1 \cap H_2, H_1 \cap L_2, L_1 \cap H_2$ and $L_1 \cap L_2$, respectively.}
\label{fig:2D_filterbank}%
\end{figure*}

Not all graphs are bipartite. In order to apply our filterbank design
to an arbitrary graph, $G = (\Vc,{\it E})$, we propose a {\em
  separable downsampling and filtering} approach, where our previously
designed two-channel
filterbanks are applied in a ``cascaded'' manner, by filtering along a series of bipartite subgraphs of the
original graph. This is illustrated in Figure~\ref{fig:2D_filterbank}. We call this a ``separable'' approach in analogy to
separable transforms for regular multidimensional signals. For example
in the case of separable transforms for 2D signals, filtering in one dimension (e.g., row-wise)
is followed by filtering of the outputs along the second dimension (column-wise). In our proposed
approach, a stage of filtering along one ``dimension'' corresponds to filtering using {\em only}
those edges that belong to the corresponding bipartite subgraph. As shown in
Figure~\ref{fig:2D_filterbank}, after filtering along one subgraph the
results are stored in the vertices, and a new transform is applied to
the resulting graph signals following the edges of the next level bipartite
subgraph. 

\begin{figure*}[htb]
\centering
\includegraphics[width=4in]{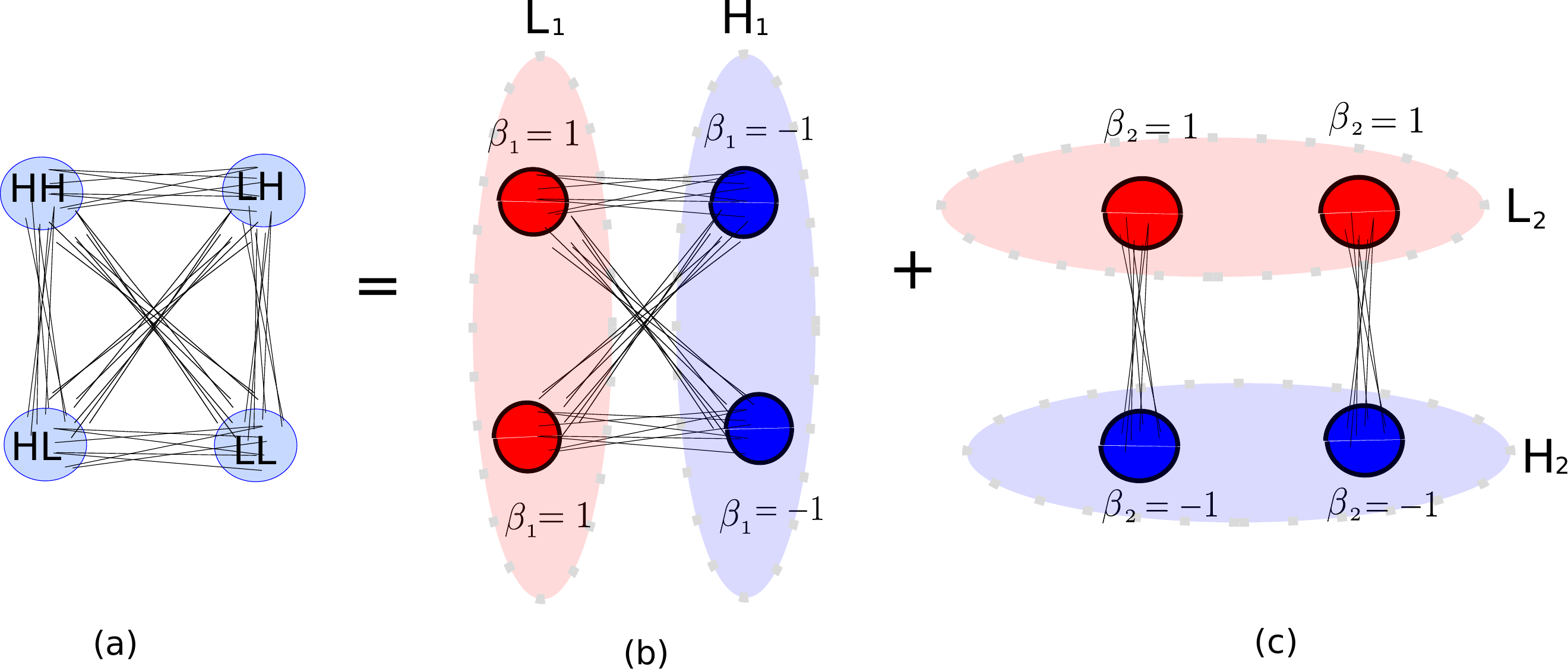}
\caption[Example of a $2$-dimensional downsampling]{\scriptsize Example of $2$-dimensional separable downsampling on a  graph: (a) original graph $G$,
 (b) the first bipartite graph $\Bc_1 = (L_1,H_1,{\it E_1})$, containing {\em all} the links in $G$ between sets $L_1$ and $H_1$. 
(c) the second bipartite graph $\Bc_2 = (L_2,H_2,{\it E_2})$, 
containing {\em all} the links  in $G - \Bc_1$, between sets $L_2$ and $H_2$}
\label{fig:2D_downsampling}%
\end{figure*}

In what follows we will assume that $G$ has been decomposed into a 
series 
of $K$ bipartite subgraphs $\Bc_i = (L_i,H_i,{\it E_i})$, $i=1 \ldots K$; how such
a decomposition may be obtained will be discussed later. 
The bipartite subgraphs cover the same vertex set: $L_i \cup
    H_i = \Vc$, $i= 1,2,...K$. Each 
edge in $G$ belongs to exactly one $E_i$, i.e., ${\it E_i} \cap {\it
  E_j} = \phi$, $i \neq j$, $\bigcup_i E_i = {\it E}$. Note that in
each bipartition we need to decide both a
2-coloring ($H_i, L_i$) {\em and} an assignment of edges ($E_i$). 
In order to guarantee invertibility for structures such as those of
Figure~\ref{fig:2D_filterbank}, 
given the chosen 2-colorings  ($H_i, L_i$), 
the edge assignment has to be performed iteratively based
on the order of the subgraphs. That is, edges for subgraph $1$ are
chosen first, then those for subgraph $2$ are selected, and so on. 
The basic idea is that at each stage $i$ {\em all} edges between vertices
of different colors {\em that have not been assigned yet} will be included
in $E_i$. 
More formally, 
at stage $i$ with sets $H_i$ and $L_i$, $E_i$ contains {\em all} the
    links in ${\it E} - \bigcup_{k=1}^{i-1} E_k$ that connect vertices
    in $L_i$ to vertices in $H_i$. 
%As an example let the first coloring be $H_1 \cup L_1 = \Vc$. 
Thus 
    $E_1$ will contain all edges between $H_1$ and $L_1$. Then, we
  will assign to $E_2$ all the links between nodes in $H_2$ and $L_2$
  that were not already in $E_1$. This is also illustrated in
  Figure~\ref{fig:2D_downsampling}. 
Note that, by construction $G_1 = G-\Bc_1 = (\Vc,E-E_1)$ contains now two
disjoint graphs, since all edges between $L_1$ and $H_1$ were assigned
to $E_1$. 
% These two disjoint graphs are denoted $G(L_1)$, $G(H_1)$,
% where 
% $G(\Sc)$, for any subset $\Sc$ of vertex set $\Vc$, 
% is the subgraph of $G$ with vertex set $\Sc$ and edge-set containing 
% all the links between the nodes in the set $\Sc$.
Thus, at the second stage in Figure~\ref{fig:2D_filterbank},  
$\Bc_2$ is composed of two disjoint graphs $\Bc_2(L_1)$ and
$\Bc_2(H_1)$, which each will be processed independently by one of the
two filterbanks
at this second stage. Clearly, this guarantees invertibility of the
decomposition of Figure~\ref{fig:2D_filterbank}, since it will be
possible to recover the signals in $\Bc_2(L_1)$ and
$\Bc_2(H_1)$ from the outputs of the 2nd stage of the
decomposition. The
same argument can be applied to the decompositions with more than two stages. That
is, the output of a two-channel filterbank at level $i$ leads to two
subgraphs, one per channel, that are disconnected when considering the
remaining edges ($E - \bigcup_{k=1}^{i} E_k$). The output of a
$K$-level decomposition leads to $2^K$ disconnected subgraphs.

We now derive expressions for the proposed cascaded transform along bipartite
subgraphs.
%  as described before to further illustrate its separability.  
Using $K=2$ case as an example, assuming that the original graph can be
approximated exactly with two bipartite subgraphs as shown in Figure~\ref{fig:2D_downsampling},
we choose $\beta_i = \beta_{H_i}$ as the downsampling function for bipartite graph $\Bc_i$, for $i= 1,2$.
Further, let us denote  
$\Jm_{\beta_{i}}$, as the downsampling matrices, and
$\Hm_{i0}$ and $\Hm_{i1}$ as the low-pass and high-pass graph-QMF filters 
respectively, for the bipartite graph $\Bc_i$, for $i = 1,2$. 
Since, the vertex sets $L_1$ and $H_1$ in bipartite graph $\Bc_2$ are disconnected,
the filtering and downsampling operations on graphs $\Bc_2(L_1)$ and $\Bc_2(H_1)$
do not interact with each other. Therefore, graph-filters $\Hm_{2j}$, for $j = 0,1$
on the second bipartite graph $\Bc_2$, can be 
represented as {\em block-diagonal matrices} with diagonal entries $\Hm_{2j}(H_1,H_1)$ and $\Hm_{2j}(L_1,L_1)$.
As a result, $\Hm_{20}$ and $\Hm_{21}$ commute 
with downsampling matrix $\Jm_{\beta_{1}}$ of the first bipartite subgraph, i.e.,
\begin{equation}
\Hm_{2j}\Jm_{\beta_{1}} = \Jm_{\beta_{1}}\Hm_{2j},
\label{eq:commute} 
\end{equation}
for $j = 1,2$.
\footnote{In general, this result can be applied to any general $K$-dimensional decomposition using proposed recursive method, as 
the downsampling matrix $\Jm_{\beta_{i}}$ commutes with all filter matrices $\Hm_{k1}$ and $\Hm_{k2}$ corresponding to bipartite subgraph $\Bc_k$, 
where $k > i$.}
 Further,  
let  $\Tm_{ai}$ be the equivalent analysis transform for $\Bc_i$, for $i = 1,2$. 
The combined analysis 
transform $\Tm_a$ in the $2$-dimensions can be written as the product of analysis transform in each dimension.  
Using (\ref{eq:analysis_tx}), we obtain:  
\begin{equation}
\Tm_a  = \Tm_{a2}.\Tm_{a1}  = \prod_{i = 1}^2 \frac{1}{2} \left((\Hm_{i1} +\Hm_{i0})  +
  \Jm_{\beta_{i}}(\Hm_{i1} - \Hm_{i0})  \right) , 
 \label{eq:multi_analysis_tx}
\end{equation}
Note that, for exact graph-QMF filter design such as with the Meyer kernel in (\ref{eq:meyer_x}), 
$\Tm_{ai}$ is invertible with $\Tm_{ai}^{-1} = \Tm_{ai}^t$, for $i= 0,1$.
As a result, $\Tm_a$ is invertible with 
$\Tm_{a}^{-1} = \Tm_{a1}^t.\Tm_{a2}^t$\footnote{For polynomial approximations, of Meyer kernels, we incur some reconstruction errors in each dimension.}. 
The transform function $\Tm_a$ can be further
% The  in (\ref{eq:multi_analysis_tx}) can be 
decomposed into the transform functions $\Tm_{HH}, \Tm_{Hl}, \Tm_{LH}$ and $\Tm_{LL}$ corresponding to the four channels 
in Figure~\ref{fig:2D_filterbank}. For example,
the transform $\Tm_{HH}$, consists of all the terms in the expansion of $\Tm_a$ in (\ref{eq:multi_analysis_tx}),  containing 
filters $H_{11}$ and  $H_{21}$.  Thus, 
\begin{equation}
\Tm_{HH} =   \frac{1}{4} (\Hm_{21}\Hm_{11} + \Hm_{21}\Jm_{\beta_1}\Hm_{11} + \Jm_{\beta_2}\Hm_{21}\Hm_{11} + \Jm_{\beta_2}\Hm_{21}\Jm_{\beta_1}\Hm_{11} ),
 \label{eq:multi_analysis_tx1}
\end{equation}
% 
% filtering and downsampling with $\Hm_{1k}$ and $(-1)^{k+1}\beta_1$ respectively on $\Bc_1$, followed 
% by filtering and downsampling with and $(-1)^{j+1}\beta_2$ respectively on $\Bc_2$.
where $(1/4)\Hm_{21}\Hm_{11}$ is the transform without downsampling, 
and the remaining terms arise primarily due to the downsampling in the $HH$ channel. 
Using (\ref{eq:commute}), which is a property of our proposed decomposition scheme in (\ref{eq:multi_analysis_tx1}), we 
% obtain:
% Since $\Bc_1$ is a bipartite graph, a 
% two-channel filterbank can be designed for $\Bc_1$, 
% using downsampling function $\beta_{H}$ 
% and the graph-QMF filters as explained in the previous Section. 
% The output coefficients 
%
obtain:
\begin{eqnarray}
\Tm_{HH} & = &  \frac{1}{4} (\Hm_{21}\Hm_{11} + \Jm_{\beta_1}\Hm_{21}\Hm_{11} + \Jm_{\beta_2}\Hm_{21}\Hm_{11} + \Jm_{\beta_2}\Jm_{\beta_1}\Hm_{21}\Hm_{11}) \nonumber \\
& = & \frac{1}{4}(\Id +\Jm_{\beta_2})(\Id + \Jm_{\beta_1})\Hm_{21}\Hm_{11}.
 \label{eq:multi_analysis_tx2}
\end{eqnarray} 
Thus, the equivalent transform in each channel of the proposed $2$-dimensional separable filterbanks can 
be interpreted as filtering with a $2$-dimensional filter, such as $\Hm_{21}\Hm_{11}$ for the $HH$ channel, 
followed by $DU$ operations with 
two downsampling functions $\beta_2(n)$ and $\beta_1(n)$ in cascade. It also follows from 
(\ref{eq:multi_analysis_tx2}), that the output of $\Hm_{21}\Hm_{11}$ in the $HH$ channel is stored 
only at the nodes corresponding to $H_1 \cap H_2$. Thus, the output of each channel is stored at 
mutually disjoint sets of nodes, and each node stores the output of exactly one of the channel. 
Therefore, the overall filterbank is {\em critically sampled}. Further, 
if the spectral decompositions of $\Bc_1$ and $\Bc_2$ are given as $\{\lambda, \Pm^1_{\lambda}\}$ and $\{\gamma, \Pm^2_{\gamma}\}$, 
then $\Hm_{21}\Hm_{11}$ consists of a two dimensional spectral kernel $h_{21}(\gamma)h_{11}(\lambda)$ and corresponding 
eigenspace $\Pm^2_{\gamma}\Pm^1_{\lambda}$. 
% This 
% is equivalent to a {\em $4$-channel decomposition} of graph-signal $\xv$, with 
% channel response $\Hm_{2j}\Hm_{1k}$ for $j,k = 0,1$ and the output coefficients stored at
% disjoint sets $LL = L_2 \cap L_1$, $LH = L_2 \cap H_1$, $HL = H_2 \cap L_1$ and $HH = H_2 \cap H_1$.
% Further, combining~(\ref{eq:commute}) and~(\ref{eq:multi_analysis_tx2}), we obtain:
% \begin{equation}
%  \Tm_{ajk}  = \frac{1}{4}(\Id + (-1)^{j+1}\Jm_{\beta_2})\Hm_{2j}(\Id + (-1)^{k+1}\Jm_{\beta_1})\Hm_{1k}
% \label{eq:multi_analysis_tx3}
% \end{equation}
% which is a separable filtering and downsampling form of the two-dimensional two-channel filterbank, 
% and is shown in Figure~\ref{fig:2D_filterbank}.
% 
% Thus, the output coefficients 
% of transform  $\Tm_{a00},\Tm_{a01},\Tm_{a10}$ and $\Tm_{a11}$.  % \begin{equation}
%  \Hm_{2j}\Hm_{1k} = \sum_{\gamma  \in \sigma(\Bc_2)}\sum_{\lambda \in \sigma(\Bc_1)}h_{2j}(\gamma)h_{1k}(\lambda)
% \label{eq:multidim_filtering}
% \end{equation}
% 
The analysis extends to any dimension $K> 2$ with $K$-dimensional 
graph-frequencies $(\lambda_{1},\lambda_{2},...,\lambda_{K})$,  
corresponding eigenspace  $\Pm_{\lambda_{1}}^1,\Pm_{\lambda_{2}}^2,...\Pm_{\lambda_{K}}^K$ and transforms with spectral response 
$\prod_{i =1}^K g_i(\lambda_{i})$.

So far we have described, how to implement separable 
multi-dimensional graph-QMF filterbanks on a graph $G$, given  
a decomposition of $G$ into $K$ bipartite subgraphs. In particular, we defined a ``separable'' method of 
graph decomposition, 
which leads to a cascaded tree-structured implementation of the multi-dimensional filterbanks.
While these multi-dimensional filterbanks can be implemented for {\em any}
separable bipartite subgraph decomposition of $G$, the definition 
of a ``good'' bipartite 
decomposition of any arbitrary graph remains a topic for future work, and may be application dependent.
%  As such, the bipartite subgraph 
% decomposition problem 
% is an active area of research in 
% graph theory and operational 
% research \cite{Harary,Plateau} and
% has applications in resource allocation and computer security. 
In this paper, we propose a bipartite subgraph 
decomposition method, referred to as {\em Harary's decomposition}, 
which provides a $\lceil log_2k \rceil$ bipartite decomposition of a graph 
$G$ given a $k$-coloring defined on it\footnote{A graph is perfectly
$k$-colorable if its vertices can be assigned 
$k$-colors in such a way that no two adjacent vertices 
share the same color. The term {\em chromatic number} 
$\chi(G)$ 
of a graph refers to smallest such $k$.}. 
The method is derived from \cite{Harary} and we describe 
it in Algorithm \ref{alg:alg1}.\footnote{Note that the bipartite decomposition is not unique and depends on the ordering in which the $k$-colors are divided.}
% The term {\em biparticity} $\theta(G)$, which refers to 
% %   of a graph $G$ is 
% the minimum number of bipartite subgraphs whose union covers 
% $G$
% % . The following 	
% % result given 
% % in \cite{Harary} provides 
% has an interesting 
% connection with 
% the {\em chromatic number}
% 
% of a graph.
% \begin{theorem}[]
% % ~\cite{Harary}
% For any graph $G$, $\theta(G) = \lceil log_2 \chi(G)\rceil$ where $\chi$ is the chromatic number of the graph and $\lceil.\rceil$ is the smallest 
% integer greater than a number. 
% \label{thm:biparticity}
% \end{theorem}
% The Lemma \ref{thm:biparticity} leads to some interesting observations. 
% For example when combined with {\em Four Color Theorem for planar graphs}, 
% it implies 
% that all planar graphs which are not bipartite can be decomposed 
% into 2 bipartite subgraphs. 
% Although the problem of determining the chromatic number $\chi(G)$ is NP-complete, there exist several approximate minimum coloring algorithms with various 
% orders of accuracies, a comparison of which can be found in~\cite{Klotz'02}. We choose a backtracking sequential coloring (BSC) 
% algorithm presented in~\cite{Klotz'02} because of its high accuracy.
% Based on this result, we propose a $\lceil log_2 k\rceil$- bipartite decomposition of the graph $G$, 
% given a perfect $k$-coloring defined on it. We refer to this method as {\em Harary's decomposition} 
% 
\begin{algorithm}[H]
 \caption{Harary's Decomposition}
 \label{alg:alg1}
 \begin{algorithmic}[1]
  \REQUIRE  $\Fm$, s.t. $F(v)$ is the color assigned to node $v$, min($F$)=$1$ , max($F$)=$k$.
  \STATE Set $L_1$ = set of nodes with $F(v) \leq \lfloor k/2 \rfloor$ colors.
  \STATE Set $H_1$ = set of nodes with $F(v) > \lceil k/2 \rceil$ colors. 
  \STATE Set $E_1 \subset E$  containing all the edges between sets $H_1$ and $L_1$. 
  \STATE Compute bipartite subgraph $\Bc_1 = (L_1,H_1,E_1)$, 
  \STATE Set $G = G - \Bc_1$. 
  \STATE $G$ is now a union of two disconnected subgraphs $G(H_1)$ and $G(L_1)$. 
  \STATE Graph $G(L_1)$ is $\lceil k/2 \rceil$-colorable. 
  \STATE Compute coloring $\Fm_L$ on $G(L_1)$ s.t. min($F_L$)=$1$ , max($F_L$)=$\lceil k/2 \rceil$.
  \STATE Graph $G(H_1)$ is $\lfloor k/2 \rfloor$-colorable. 
  \STATE Compute coloring $\Fm_H$ on $G(H_1)$ s.t. min($F_H$)=$1$ , max($F_L$)=$\lfloor k/2 \rfloor$.
  \STATE Repeat $1-4$ on $G(L_1)$ and $G(H_1)$ to obtain bipartite subgraphs $\Bc_{2}(L_1)$ and $\Bc_{2}(H_1)$.
  \STATE Compute bipartite subgraph $\Bc_2 = \Bc_{2}(L_1) \cup \Bc_{2}(H_1)$.
  \STATE Set $G = G - \Bc_2$. 
  \STATE repeat $1-13$ exactly $\lceil log_2 k\rceil$ times after which graph $G$ will become an empty graph. 
 \end{algorithmic}
\end{algorithm}

% Text for non-separable solution

Note that invertible cascaded transforms can also be constructed even
when the conditions for edge selection described are not
followed, e.g., if an edge $e_1$ between nodes in $H_1$ and $L_1$ is {\em not}
included in $E_1$. In such a situation, it is possible to perform an
invertible cascaded decomposition if $e_1$ is no longer used in
further stages of decomposition. Thus,
we would have an invertible decomposition but on a graph that
approximates the original one (i.e., without considering
$e_1$). Alternatively it can be shown that it is possible to design
invertible transforms with arbitrary $E_i$ selections (i.e., not
following the rules set out in this paper), but these transforms are
not necessarily critically sampled. A more detailed study of this case
falls outside of the scope of this paper.  
\add[SN]{\protect \subsection{Multiresolution decomposition using two-channel filterbanks}
The two-channel filterbanks on a single  bipartite graph $\Bc = (H,L,{\it E})$  
have the property 
of decomposing the signal into two lower-resolution versions $\hat \fv_{L}$ and $\hat \fv_{H}$ respectively, as in (\ref{eq:channel_outp}). 
The signal $\hat \fv_{L}$ is a lowpass or 
coarse resolution version constructed from the output coefficients of the lowpass channel stored on the set $L$, whereas
$\hat \fv_{H}$ is a highpass version of the input constructed from the output coefficients of 
the filterbank stored on the set $H$. Analogous to tree-structured filterbanks for $1$-D signals, 
this decomposition can be applied recursively on the low-pass (or high-pass) signal by constructing a downsampled graph 
consisting of vertices in $L$ (or $H$) and some appropriate edge-structure.  One way to 
compute the downsampled graph $G_L$ (or $G_H$) is to reconnect two nodes in set $L$ (or $H$) if 
they are $2$-hops away in the original graph. 
% While the downsampled graphs obtained using this method have a well known 
% interpretation for regularly sampled signal domains, the extension to bipartite graphs is not known.  
Note that for bipartite graphs, unlike the case of regular lattices,
the resulting downsampled graphs $G_L$ and $G_H$ may neither be identical nor bipartite. Therefore, 
for the next level of decomposition, we can either operate on a single bipartite graph approximation of $G_L$ which leads to a 
one-dimensional two-channel filterbank, or a multiple bipartite graph approximation, which leads to a multi-dimensional 
two-channel filterbank implementation on the downsampled graph. Further, this multiresolution 
decomposition of graph-signals can be extended to the case of  general $K$-dimensional 
two-channel filterbanks for any arbitrary graph $G$, which decomposes the signal into 
$2^K$ lower-resolution versions, as described in Section \ref{sec:bpt_decomp}. In this case, the downsampled 
graphs in each channel, can be computed by reconnecting two nodes in the downsampled vertex set, if 
they are $2^K$-hops away in the original graph.
% This leads to a hierarchy of resolutions which can be termed as a 
% .
}
\section{Experiments}
\label{sec:experiments}
\subsection{Graph-QMF Design Details}
\label{sec:fb_details}
We first provide 
explicit details of the filterbank design for 
arbitrary graphs. 
Given any arbitrary 
undirected graph $G= (\Vc,E)$,
we find a minimum perfect-coloring $\chi$ 
of its vertices using a graph-coloring 
algorithm,  
such as the 
BSC algorithm given in~\cite{Klotz'02}. 
\change[SN]{\protect This is then 
used in Harary's Algorithm~\ref{alg:alg1} to find set of  $K = \lceil log_2(\chi)\rceil$ disjoint bipartite graphs 
$G_i = (L_i,H_i,E_i),i=\{1,2,...K\}$.}{\protect The coloring information is then used to decompose $G$ into 
a set of $K = \lceil log_2(\chi)\rceil$ bipartite graphs 
$\Bc_i = (L_i,H_i,E_i)$, for $i=1,2,...K$ 
using Harary's Algorithm as described in Section~\ref{sec:bpt_decomp}.
}
For each subgraph  $\Bc_i$, 
% a wavelet filterbank is constructed on each bipartite subgraph as follows:
%  For each bipartite graph $G_i$ 
we compute its normalized 
Laplacian matrix $\Lcb_i$ and the downsampling function $\beta_i = \beta_{H_i}$.  
Further, we compute the  
low-pass analysis kernel $h_{i,0}(\lambda)$ on $\Bc_i$, 
as the $m_i^{th}$ order 
Chebychev approximation of the Meyer kernel $h_0^{Meyer}(\lambda)$, for some positive integer value $m_i$.
The remaining spectral 
kernels $h_{i,1}(\lambda),g_{i,0}(\lambda),g_{i,1}(\lambda)$ are computed from $h_{i,0}(\lambda)$ according 
to graph-QMF relations mentioned in (\ref{eq:graph-QMF}).
 The corresponding analysis and synthesis transforms are 
then computed as $\Hm_{i,j} = h_{j}(\Lcb_i)$ and $\Gm_{i,j} = g_{j}(\Lcb_i)$, respectively, for $j =0,1$. 
Note that, since the kernels are polynomials, 
the transforms are 
also matrix polynomials of Laplacian matrices 
and do not 
require explicit eigenspace 
decompositions. 
In our experiments, we use $m_i = m$, and 
hence $h_{i,j}(\lambda) = h_j(\lambda),~j \in \{0,1\}$ for all $i$,
%  for a simple design,
in which case the resulting transforms 
are exactly $m$-hop 
localized on each bipartite subgraph. 
The order $m$ is a 
parameter of our design and should be chosen 
based on the required level of spatial localization and how 
much reconstruction error can be tolerated.
% below desired level in each subgraph $G_i$. 
The overall filterbank is designed by 
concatenating filterbanks of each 
bipartite subgraph in the form of a tree, analogous to Figure~\ref{fig:2D_filterbank} in the $2$-dimensional 
decomposition case.   
We now describe some experiments to demonstrate potential applications
of our proposed filterbanks. 
% (blue plots).  
% The Meyer wavelet approximations follow the constraint 
% $h_0(\lambda)^2 + h_0(2 - \lambda)^2 - c^2 = 0$  
% more closely than ideal-filter approximation. This is shown in Figure~\ref{fig:ideal_filt_approx_constraint}. 
% % the constraint 
% % function in both cases. 
% % Ideally the constraint function should be zero. 
% \begin{figure}[htb]
% \centering
% \includegraphics[width=7in]{figs/kernel_approx_constraints.eps}
% \caption[A 2D image with 8-connectivity decomposed into two bipartite subgraphs.]{\footnotesize
% The constraint plots}
% \label{fig:ideal_filt_approx_constraint}%
% \end{figure}
% However for 
% polynomial approximations  the constraint function have variations. 

\subsection{Graph Filter-banks on Images}
\begin{figure*}[htb]
\centering
\includegraphics[width=6.5in]{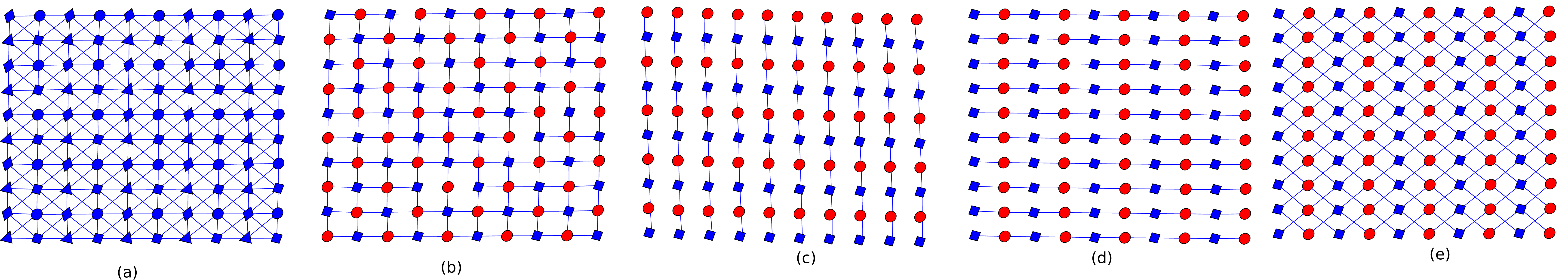}
\caption[Some of the graph-formulation of a 2D image ]{\footnotesize
Some of the graph-formulation of a 2D image lattice:
(a) shows an $8$-connected image graph $G$ formed by connecting each pixel with its $8$ nearest neighbors. 
The graph is $4$-colorable, and the nodes of different shapes 
(squares,circles,triangles and diamonds) represent different colors. 
(b) shows the image-graph $G^r$ by connecting each pixel with its rectangular (NWSE) neighbors only, 
(c) the image graph $G^v$ 
with vertical links only 
(d) the image-graph $G^h$ with horizontal links only. 
and (e)  shows image-graph $G^d$ with each pixel linked to its $4$ diagonal neighbors  
The graphs shown in (b), (c), (d) and (e) are bipartite graphs,  
with the partitions represented as nodes with different 
colors and shapes (red-circles vs. blue-squares).
% 
% with .  Graph $G$ can be decomposed into a rectangular bipartite 
% subgraph $G^r$ and a diagonal bipartite subgraph $G^d$. 
% The bipartitions in the subgraphs are represented by red and blue colors.
}
\label{fig:8connected_decomp1}%
\end{figure*}

Digital images are $2$-D regular signals, but  
they can also be formulated as 
graphs by connecting every pixel (node) 
in an image with its 
neighboring pixels (nodes) and by interpreting 
pixel values as the values of the 
graph-signal at each node. The 
graph-representations of the regular-signals are shown to 
be promising in practice recently~\cite{shen2010edge,martinezvideo}.
Figure~\ref{fig:8connected_decomp1} 
shows some of the ways in which pixels in an image 
can 
be connected with each other 
to formulate a graph representation 
of 
any image.
The advantage of using a
graph 
formulation 
of the images is 
that it provides
flexibility of linking 
pixels in 
arbitrary ways, leading to 
different filtering/downsampling 
patterns. 
% What we observe is that,
% Many known downsampling schemes 
% such as rectangular downsampling, 
% quincunx downsampling etc.,
% can be  explained
% % understood 
% in terms of 
% downsampling  
% % in 
% % corresponding 
% bipartite image graphs with particular  
% connectivities. 
\begin{figure*}[htb]
\centering
\includegraphics[width=6.5in]{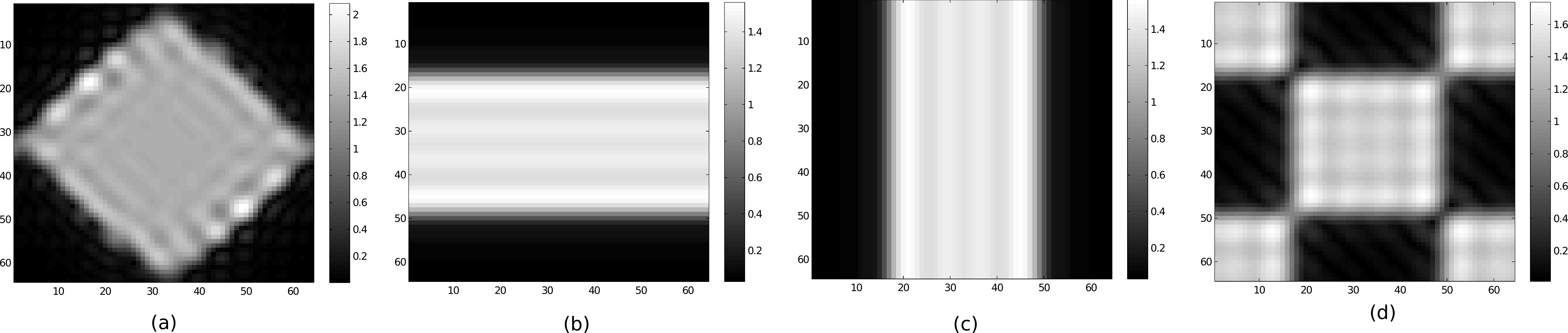}
\caption[Fourier frequency responses of ideal lowpass filters.]{\footnotesize Discrete Fourier frequency magnitude 
responses of ideal lowpass filters on some bipartite image-graphs.
Fig. (a) ideal lowpass filter response on NWSE bipartite subgraph $G^r$ shown in \ref{fig:8connected_decomp1}b,
Fig. (b) ideal lowpass filter response on diagonally connected bipartite subgraph $G^d$ shown in \ref{fig:8connected_decomp1}c, 
Fig. (c) ideal lowpass filter on vertical-links only bipartite subgraph shown in  \ref{fig:8connected_decomp1}d,  
Fig. (d) ideal lowpass filter on horizontal-links only bipartite subgraph shown in  \ref{fig:8connected_decomp1}e. }
\label{fig:spect_freq}%
\end{figure*}
% 
% different ways 
% %  in the image. 
% % Besides, 
% 
% which and 
% hence changing the 
% orientation 
% A lot of 
% research has been done to 
% design $2$-dim filterbanks 
% which can process 
% these images. 
% 
% A simple way of designing these filterbanks 
% is to design separable $1$-dim 
% filters along mulitple directions (orientaions) (eg. horizontal 
% and vertical).
% We first demonstrate that our proposed graph based 
% filterbanks 
% can handle multi-dimensional regularly sampled 
% signals such as 
% digital images  
% and provide 
% % results which have 
% similar 
% interpretation as 
% % that of 
% classical filterbanks. 
% % to form . 
% 
%  However this is not an 
% efficient transform for images whose 
% edges are 
% not aligned along these directions. 
% To overcome this we 
% can choose additional 
% direction (eg. directional DCTs) to filter/downsample the signal which 
% however leads to an oversampled transform. 
% research is 
% 
% The filterbank designs on images
% 
% of pixels each representing a color 
% or an intensity value.  The lattice is represented by a $2\times 2$ nonsingular integer matrix 
% $\Mm$ as 
% \begin{equation}
%  
% \end{equation}
% 
% The theory of two-dimensional filter-banks is well known to 
% signal processing community.  
% 
% 
% The filter-banks designs suc
% We present an alternative way of designing 
% filterbanks on images which provide 
% directional filtering 
% while still being critically sampled. 
% Digital images 
% % are 2D signals 
% % which 
% can be interpreted as 
% % a 
% two-dimensional 
% regular lattice graphs by 
% In this 
To demonstrate this, we implement an
ideal spectral low-pass 
filters on the graph formulations of the $2$D images, shown 
in Figure~\ref{fig:8connected_decomp1}. 
Since, the graphs $G^r,G^v,G^h$ and $G^d$ are all bipartite graphs, 
the ideal spectral lowpass filter $\Hm_0^{ideal}$ on these graph can be 
computed as in~(\ref{eq:ideal_filt}). 
% Note that, $\Hm_0^{ideal}$ 
% % 
% %  and
% is also   
% % acts as 
% the anti-aliasing filter on the graph because of 
% the spectral folding phenomenon. 
% Further, the anti-aliasing filter for any 
% arbitrary graph is the product of ideal-lowpass filters along its bipartite subgraph decompositions.
In Figure~\ref{fig:spect_freq}, 
we plot the {\em DFT} magnitude response 
of ideal lowpass spectral 
transforms on bipartite image-graphs 
$G^r,G^v,G^h$ and $G^d$ respectively.\footnote[5]{Because of the 
regularity and symmetry of the links, 
the resulting filters at each node, 
% i.e., the rows of transform matrices $\Hm_i$, 
are translated version of each other
(except at the boundary nodes), 
and so we can compute the 
$2$-D
DFT magnitude response of a spectral transform, by computing 
the DFT response of the filtering operations at a single node.} 
%  on 
% bipartite subgraphs $G^r$ and $G^d$ 
% and the original graph $G$.
% ,  for bipartite 
% graphs or proposed bipartite 
% subgraph decompositions 
% of a graph. 
% 
% we construct an image graph by 
% connecting every pixel
% with its 
% % $4$ NWSE or $4$ diagonal neighbors or 
% % more generally all 
% $8$ nearest 
% neighbors ($4$ NWSE or $4$ diagonal) as shown in Figure~\ref{fig:8connected_decomp1}(a). 
% % An image can be interpreted as a graph  
% The $8$ connected image-graph 
% % In a more general 
% % analysis we connect every node with all of its
% % . 
% % The resulting image graph 
% $G$ is planar, i.e., the edges do not cross-over and 
% % but is not bipartite. 
% % It can be seen in , 
% % As seen in Figure \ref{fig:8connected_decomp1}(a), 
% % The 
% its
% chromaticity is $\chi = 4$ (represented as different shape nodes in  Figure~\ref{fig:8connected_decomp1}(a)). 
% % of an $8$-connected regular lattice 
% % is four. and 
% By Theorem~\ref{thm:biparticity}, it can be decomposed into two disjoint 
% bipartite subgraphs and from several such possible 
% decompositions, we 
% choose the decomposition which gives 
% us a rectangular subgraph $G^r$ and a 
% diagonally connected diamond graph $G^d$ 
% as shown in 
% Figure~\ref{fig:8connected_decomp1}(b)-(c). 
% The bipartitions in each subgraphs are represented 
% as different colored (red/blue) nodes.
In Figure
\ref{fig:8connected_decomp1}(b) the downsampling 
pattern (red/blue nodes) on the 
rectangular subgraph $G^r$ is identical to
the quincunx downsampling pattern, and in 
Figure~\ref{fig:spect_freq}(a), it can be observed 
that the DFT magnitude response of the spectral low-pass 
filter on $G^r$
is same as the 
DFT magnitude response of the standard anti-aliasing 
filter for quincunx downsampling. Similarly, 
we observe that the spectral low-pass filters for 
$G^v$ in Figure \ref{fig:8connected_decomp1}(c) and $G^h$ in Figure \ref{fig:8connected_decomp1}(d) have the 
same DFT magnitude responses (Figure~\ref{fig:spect_freq}(b) and~\ref{fig:spect_freq}(c))
as the anti-aliasing filters for vertical and horizontal factor-of-$2$ downsampling 
cases, respectively.
Further, the graph formulation of images allows 
us to explore new downsampling patters, for example, the image pixels 
can be connected to 
their diagonally opposite neighbors as shown in Figure \ref{fig:8connected_decomp1}(e). 
The DFT magnitude response of the ideal spectral low-pass filter in this case, 
is shown in Figure~\ref{fig:spect_freq}(d) and 
has a wider passband in the diagonal directions. 
% This makes sense as the pixel connectivity
% is oriented in the diagonal directions. 
Further, in the non-bipartite graph formulation of the anti-aliasing filter for any 
arbitrary graph is the product of 
ideal-lowpass filters along its bipartite subgraph decompositions. Therefore 
the rectangular 
graph $G^r$ can be further decomposed into bipartite subgraph $G^v$ and $G^h$
% a vertical-links only subgraph and a horizontal-links only subgraph, 
leading to a rectangular (factor of 4) downsampling pattern. 

\begin{figure*}[htb]
\centering
\includegraphics[width=5in]{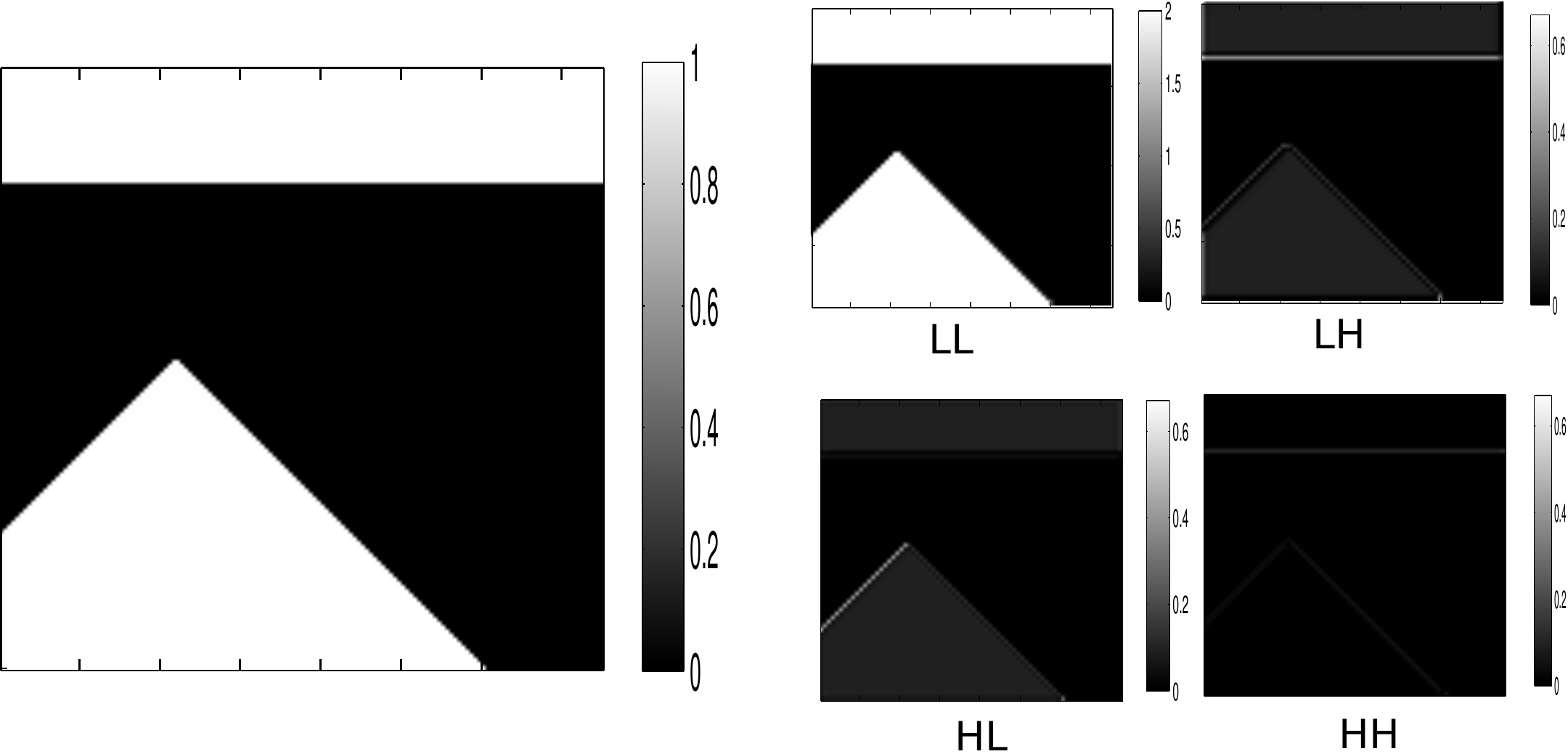}
\caption[Filterbank on Images]{\footnotesize Separable two-dim 
two channel graph filterbank on a toy image with both rectangular 
and diagonal edges. The filterbank is the concatenation of proposed graph-QMF filterbank with $m=2$ 
order approximation of Meyer kernel on subgraph $G^r$ and subgraph $G^d$ 
as shown in Figure~\ref{fig:2D_filterbank}}
\label{fig:single_bpt_images}%
\end{figure*}

\begin{figure*}[htb]
\centering
\includegraphics[width=5in]{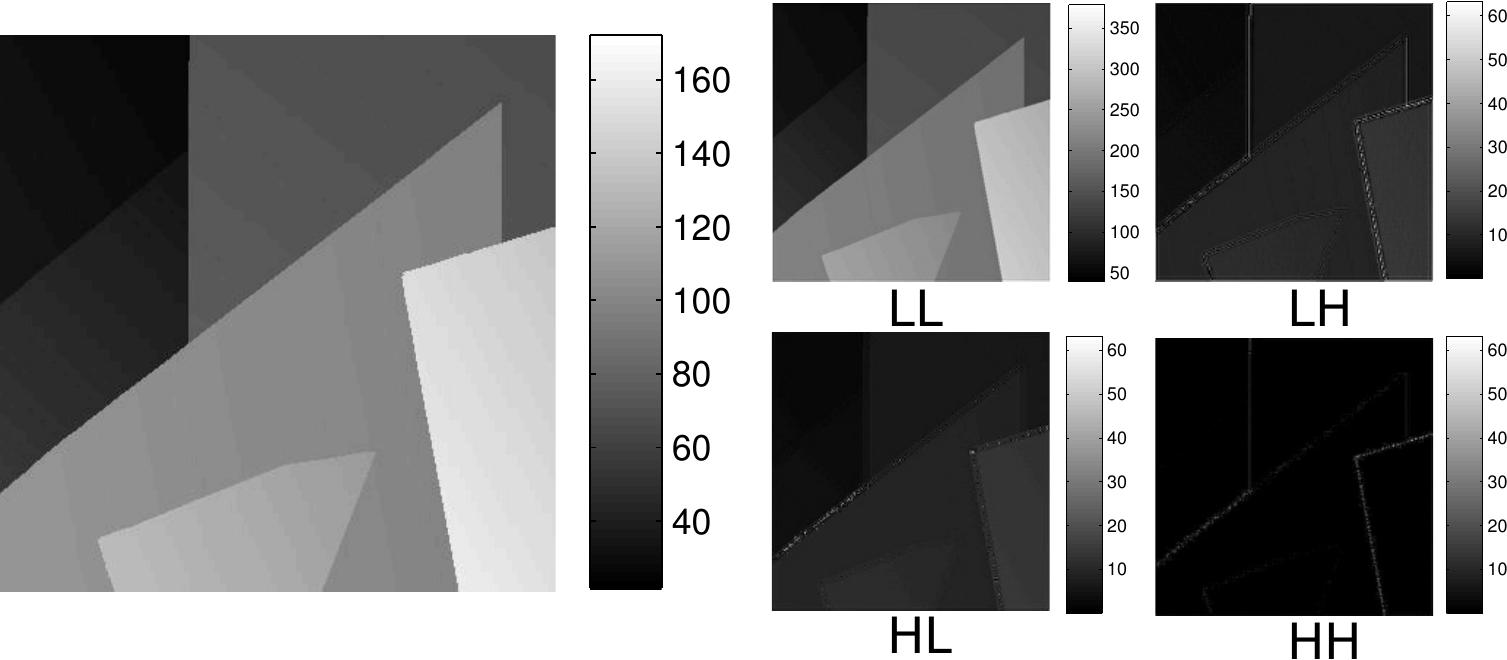}
\caption[Example of an $8$-connected image]{\footnotesize 
Separable two-dim 
two channel graph filterbank on a depth-map image with parameter $m = 6$}
\label{fig:double_bpt_images}%
\end{figure*}

This graph-based approach also 
provides additional degrees of freedom (directions)
to filter/downsample the image 
while still having a critically sampled output.
% Moreover, since 
% graph-based transforms operate only over 
% the links between nodes, the graph 
% formulation 
% % can also be 
% is useful in 
% designing edge-aware 
% transforms (which avoid filtering 
% across edges) by removing links between 
% pixels across edges. 
To demonstrate this, we implement a graph wavelet filterbank on 
the $8$-connected image-graph $G$ of a given image. The
% as shown in Figure~\ref{fig:8connected_decomp1}. 
% The 
chromaticity of $G$
% of $G$ 
is $\chi = 4$ (represented as different shape nodes in  
Figure~\ref{fig:8connected_decomp1}(a)) and hence 
% of an $8$-connected regular lattice 
% is four. and 
it can be decomposed into two edge-disjoint 
bipartite subgraphs. Among several such possible 
decompositions, we choose the decomposition that gives 
us a rectangular subgraph $G^r$ and a 
diagonally connected diamond graph $G^d$. On each subgraph we 
implement a graph-QMF filterbank, as described in 
Section~\ref{sec:fb_details} above. % with 
% % $h(\lambda)$ as 
% the $m=2$ order polynomial 
% kernel approximation of Meyer kernel $h_0^{meyer}(\lambda)$ 
% . 
The resulting 2-dim separable filterbank has four channels as 
shown in Figure~\ref{fig:2D_filterbank} and the 
% downsampling pattern 
% is such that 
nodes representing a specific shape in Figure~\ref{fig:8connected_decomp1}(a) store the 
output of a specific channel.\footnote[6]{In general for an arbitrary graph with $K$-proper colors, the bipartite decomposition provides exactly 
$K$ non-empty channels and nodes of a particular color store the output of a particular channel.} 
Figure~\ref{fig:single_bpt_images} shows the output wavelet coefficients of proposed $2$-dim filterbank on 
a toy image which has both diagonal and rectangular edges. In Figure~\ref{fig:single_bpt_images}, 
the energy of wavelet coefficients in the LH channel 
(low-pass on $G^r$, high-pass on $G^d$) is high around the rectangular edges, 
which is reasonable, since subgraph $G^d$ is diagonally connected and its low-pass spectral frequencies
are oriented along diagonal links. Similarly we observe that the high-energy wavelet coefficients 
in the HL channel 
(high-pass on $G^r$, low-pass on $G^d$) lie around the diagonal edges, since $G^r$ is 
rectangularly connected and its low-pass spectral frequencies are oriented towards horizontal 
and vertical directions. This example also shows that the filterbanks based 
on only NWSE connectivity are 
more suited for images with 
horizontal and vertical edges whereas 
the transform based only on diamond connectivity 
are 
more suited for image with diagonal edges. In Figure~\ref{fig:double_bpt_images}, 
we show the graph-wavelet decomposition of a depth-map image taken from~\cite{depth_map_dataset'01}. Again, 
we see that the LH channel has high energy coefficients along nearly rectangular edges while the HL 
channel has high energy coefficients along nearly diagonal directions. More directions can 
be added to downsample/filter by increasing the connectivity of the pixels in the image-graph. 
Moreover, since 
graph-based transforms operate only over 
the links between nodes, the graph 
formulation 
% can also be 
is useful in 
designing edge-aware 
transforms, such as \cite{shen2010edge,martinezvideo}, (which avoid filtering 
across edges) by removing links between 
pixels across edges.

\subsection{Graph Filter-banks on Irregular Graphs}
\begin{figure*}[htb]
\centering
\includegraphics[width=4.5in]{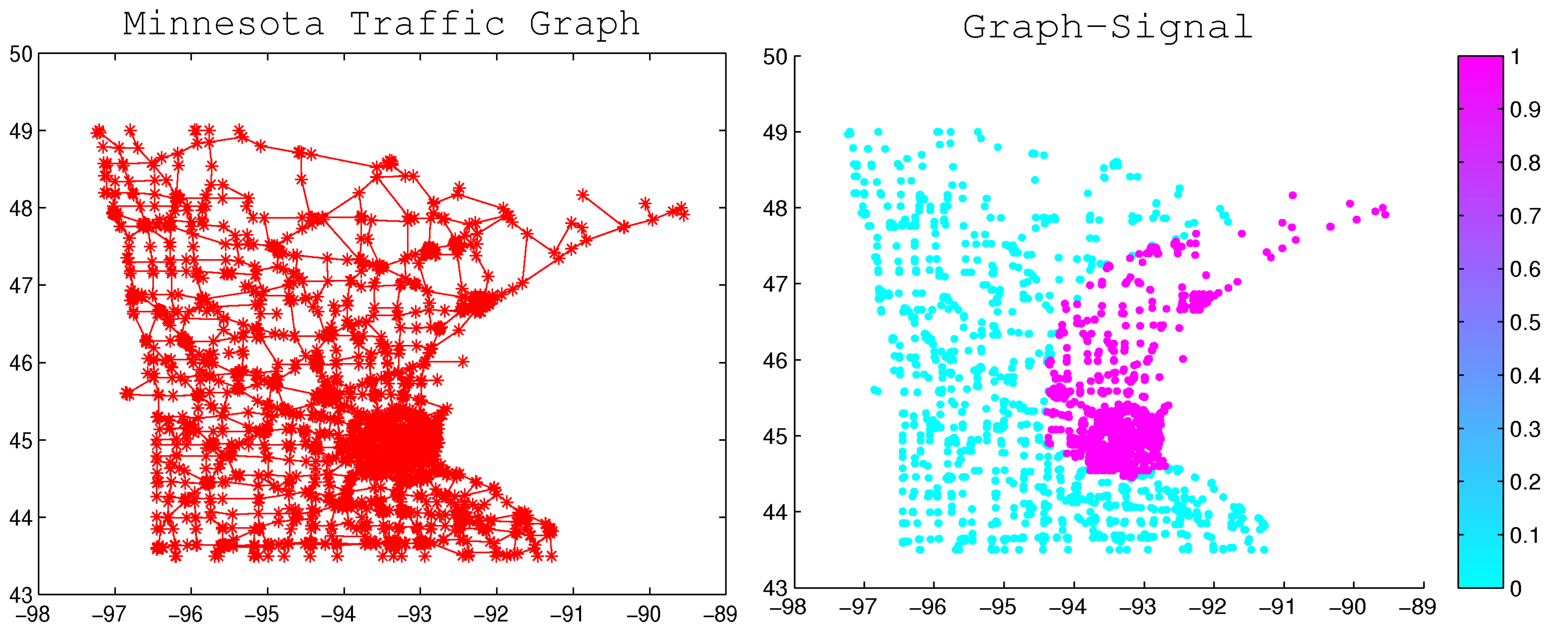}
\caption[inputs]{\footnotesize 
(a) The Minnesota traffic graph and (b) the scatter-plot 
of a graph-signal to be analyzed. The colors of the nodes represent the sample values. 
% For simplicity 
% we operate on binarized values of this signal.
}
\label{fig:minnesota_inp}%
\end{figure*}

\begin{figure*}[htb]
\centering
\includegraphics[width=4.5in]{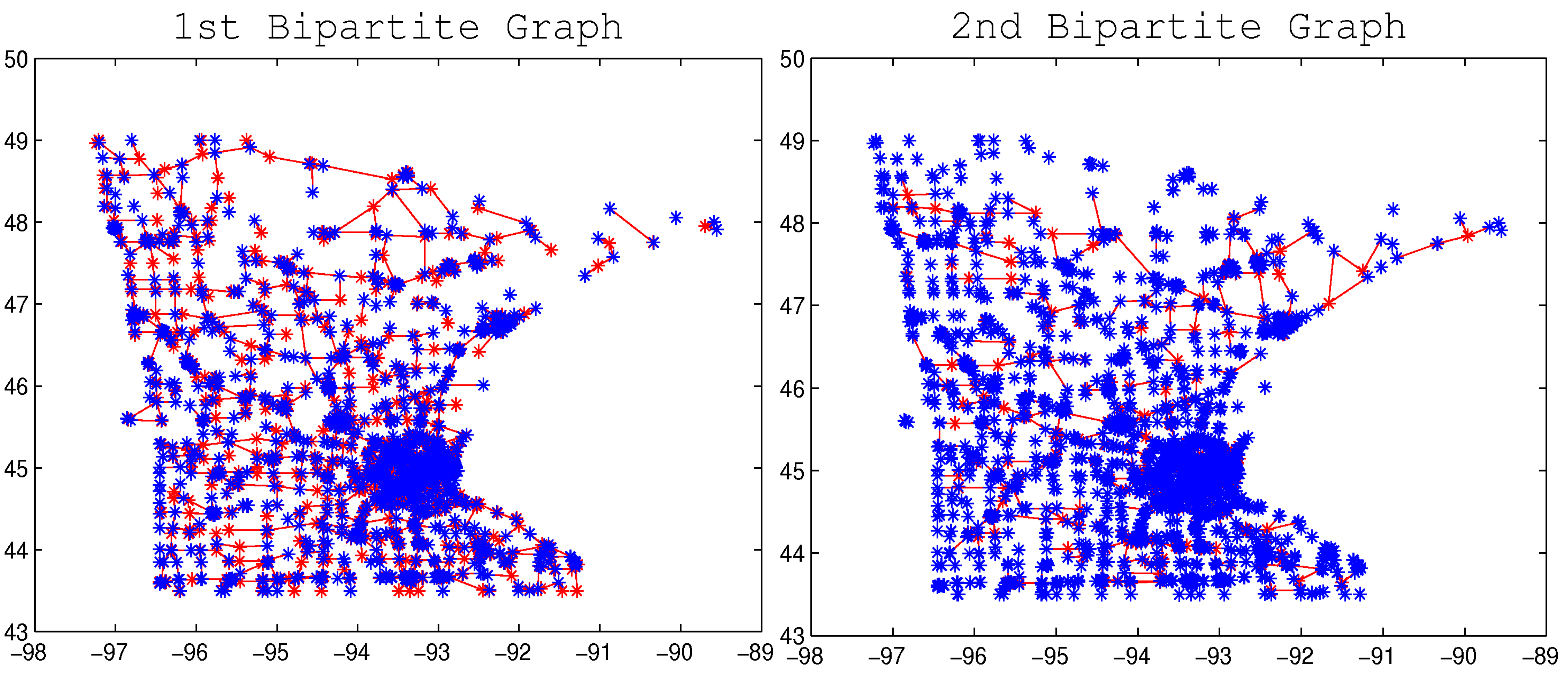}
\caption[inputs]{\footnotesize 
Bipartite decomposition of Minnesota graph into two bipartite subgraphs using Harary's decomposition.
% For simplicity 
% we operate on binarized values of this signal.
}
\label{fig:minnesota_bpt_decomp}%
\end{figure*}

Our proposed filterbanks can be used as a useful tool in analyzing/compressing 
arbitrarily linked irregular graphs. 
In order to demonstrate it we take the example of
{\em Minnesota traffic graph} $G$ as shown in Figure~\ref{fig:minnesota_inp}(a). 
Further, we consider the decomposition of a graph-signal 
whose scatter-plot is shown in 
Figure~\ref{fig:minnesota_inp}(b), 
using our proposed filterbanks on graph.  
% 
% % there are 
% % $3$ possible such decomposition and we choose the 
% % one which provides the most balanced 
% % partitioning of links in the two subgraphs. 
% we analyze a binary-value function on the graph 
% 
% Suppose, we want to analyze a graph-signal 
The graph
is perfectly $3$-colorable and hence, we can decompose it 
into $\lceil log_2(3) \rceil = 2$ bipartite subgraphs $\Bc_1$ and $\Bc_2$ 
which are shown in 
Figure~\ref{fig:minnesota_bpt_decomp}(a) and~\ref{fig:minnesota_bpt_decomp}(b) 
respectively.  

\begin{figure*}[htb]
\centering
\includegraphics[width=4.5in]{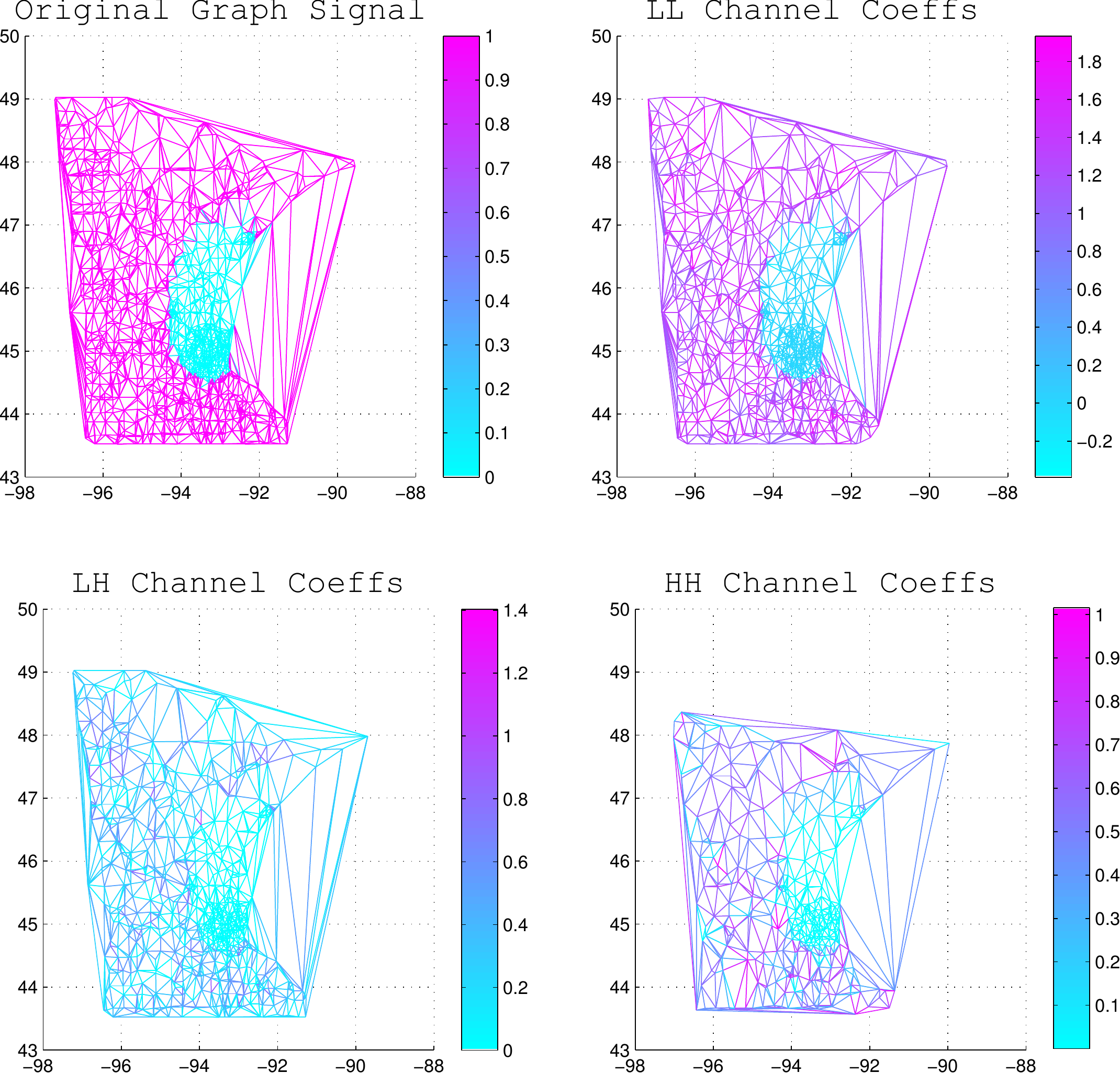}
\caption[Channel Ouputs]{\footnotesize 
The Delaunay triangulation plots of output wavelet coefficients 
of the
proposed filterbanks with parameter $m = 6$. The edge-color reflects the value of the coefficients at that point.  
(a) original graph signal 
(b) LL channel wavelet coefficients 
(c) LH  channel wavelet coefficients (d) HH channel wavelet coefficients }
\label{fig:minnesota_outp}%
\end{figure*}

\begin{figure*}[htb]
\centering
\includegraphics[width=4.5in]{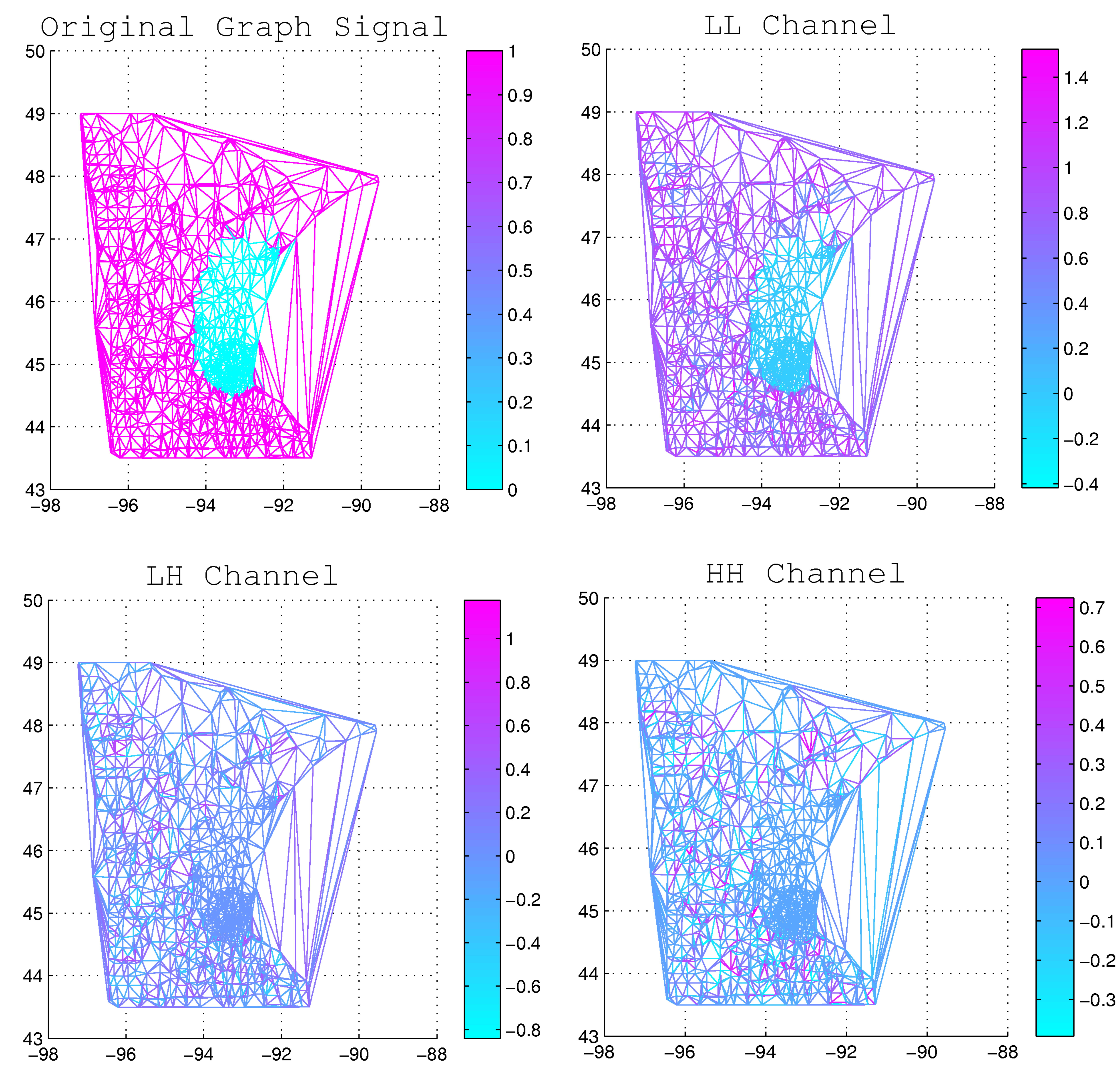}
\caption[Reconstructed image]{\footnotesize 
The Delaunay triangulation plots of the reconstructed graph-signals using the coefficients of a single channel. As before 
the edge-color reflects the value of the coefficients at that point.   (a) original graph-signal 
(b) reconstruction with LL channel coefficients only (c) reconstruction with LH channel coefficients only  
(d) reconstruction with HH channel coefficients only}
\label{fig:minnesota_reconst}%
\end{figure*}

Given such decomposition, we implement a $2$-dim separable 
$2$-channel graph-QMF filterbank 
on the Minnesota graph, with 
parameter value $m=6$, according to the details given in Section~\ref{sec:fb_details}.
% Note that 
% due to Lemma~\ref{lem:polynomial_kernel} the resulting 
% transform is $6$-hop localized on 
% each bipartite subgraph. 
Since the proper coloring of graph $G$ is $3$, there are no nodes 
to sample $HL$ channel output (i.e. nodes for which $(\beta_1(n),\beta_2(n)) = (-1,1)$) and hence
there are 
only three non-empty channels ($LL,LH,HH$). 
% Similar to the case of image-graph, 
% the downsampling in each 
% bipartite subgraph is governed by binary downsampling 
% functions ( respectively) 
% and each node $n$ in the original graph store the outcome of only one of the 
% channel (lowpass-lowpass,lowpass-highpass,highpass-lowpass and highpass-highpass) 
% depending on the state of $(\beta_1(n),\beta_2(n))$. It turns out that since there 
% is no node $n$ in the original graph with $(\beta_1(n),\beta_2(n)) = (-1,1)$, one of 
% the channel (highpass-lowpass) 
% Thus using 
% our algorithm number of non-empty channels are equal to total number of unique perfect colors 
% assigned to the nodes.
Figure~\ref{fig:minnesota_outp} shows 
the output wavelet coefficients. The $HL$ channel is empty and is not displayed in the results. 
Due to downsampling, the total number of 
output coefficients in the four channel is equal to number of input samples, thus 
making the transform {\em critically sampled}. We observe in Figure~\ref{fig:minnesota_outp} 
that for the $LL$ channel $(\beta_1(n),\beta_2(n)) = (1,1)$, 
the signal on the downsampled graph is a smooth approximation of the original signal 
(sharp boundaries blurred).
% resulting graph is a downsampled version of the original graph. 
% Further the 
% output  on the downsampled graph is a lowpass approximation of the original data . 
The remaining channels store the detail information required to perfectly reconstruction, 
original graph signal from its smooth approximation.
% For better visualization, 
% we do a hard thresholding of the detail coefficients and only show the top $40 \%$ values. 
In order to see how much energy of the original signal is captured in each channel, 
we upsample then filter the coefficient 
of each channel by the synthesis part of proposed filterbank. Figure~\ref{fig:minnesota_reconst} 
shows the output 
of each of the four channel after upsampling/filtering. We see in these plots, that 
Figure~\ref{fig:minnesota_reconst}(b) is 
an approximation of the original signal, 
while Figure~\ref{fig:minnesota_reconst}(c), and Figure~\ref{fig:minnesota_reconst}(d) are the details required to 
reconstruct the original signal from the approximation. 
% \subsection{Graph Compression}
% \subsection{Social Network Analysis}
\section{Conclusion and Future Work}
\label{sec:conclusions}
We have proposed the construction of critically sampled 
% a framework for designing two-channel 
wavelet filterbanks
% perfect reconstruction wavelet-like filterbanks 
for analyzing graph-signals
% functions 
defined on any
arbitrary finite weighted graph. For this we have formulated 
a bipartite subgraph 
decomposition problem 
which produces an edge-disjoint collection of bipartite 
subgraphs. For these bipartite graphs we have described and proved 
a {\em spectrum  folding} phenomenon which occurs in
downsampling then upsampling ($DU$) operations and produces {\em aliasing} 
in the graph signals. Based on this result, 
we have proposed two-channel wavelet 
filterbanks on bipartite graphs 
and 
% have 
provided necessary and sufficient 
conditions for aliasing 
cancellation, 
perfect reconstruction 
and orthogonality
in these filterbanks. As a practical solution, 
% 
% We show that  
% a {\em spectral folding} phenomenon in {\em bipartite graphs} occurs 
% during   Especially  
we have proposed a graph-QMF design for bipartite graphs which 
has all the above mentioned features. 
The filterbanks 
are however, realized 
by Chebychev polynomial approximations at the cost of small reconstruction error 
and loss of orthogonality. Our current efforts are 
focused on finding solutions other than the proposed graph-QMF design 
and to understand and differentiate `good' and 'bad' decompositions 
of arbitrary graphs into 
bipartite subgraphs. 
% for more meaningful results. 
\section{Acknowledgements}
The authors would like to thank Dr. David Shuman at EPFL, for his comments on 
the first version of the manuscript. We would also like to thank 
the anonymous reviewers and the associate editor,
for their valuable comments and suggestions to improve the
quality of the paper. 
\bibliographystyle{IEEEbib}
\bibliography{refs}
\end{document}